\documentclass[floatfix,showpacs,amsmath,amsfonts,amssymb,aps,
twocolumn,superscriptaddress,pra,10pt]{revtex4-1}
\usepackage[pdftex]{graphicx}

\newcommand{\dd}{\mathrm{d}}
\newcommand{\vect}[1]{\mathbf{#1}}
\newcommand{\kf}[0]{k_\mathrm{F}}
\newcommand{\Ef}[0]{E_\mathrm{F}}
\newcommand{\rc}[0]{r_\mathrm{c}}

\newcommand{\expe}[0]{\mathrm{e}}
\newcommand{\Ec}[0]{E_\mathrm{c}}
\newcommand{\kc}[0]{k_\mathrm{c}}
\newcommand{\figref}[1]{Fig.~\ref{#1}}

\newcommand{\appref}[1]{Appendix~\ref{#1}}

\newcommand{\punc}[1]{\,#1}
\newcommand{\neweqnline}{\nonumber\\}
\newcommand{\secref}[1]{Section~\ref{#1}}
\newcommand{\eqnref}[1]{Equation~(\ref{#1})}

\newcommand{\dbar}{\mathchar'26\mkern-9mu d}
\newcommand{\omegaiso}[0]{\omega_\mathrm{iso}}

\newcommand{\thetac}[0]{\theta_\mathrm{c}}
\newcommand{\bigRs}[0]{R_\mathrm{s}}

\begin{document}

\title{Pseudopotentials for an ultracold dipolar gas}
\author{T.M.~Whitehead}
\affiliation{Cavendish~Laboratory, J.J.~Thomson~Avenue, Cambridge, CB3~0HE, 
United Kingdom}
\author{G.J.~Conduit}
\affiliation{Cavendish~Laboratory, J.J.~Thomson~Avenue, Cambridge, CB3~0HE, 
United Kingdom}
\date{\today}

\begin{abstract}
A gas of ultracold molecules interacting via the long-range dipolar potential 
offers a highly controlled environment
	in which to study strongly correlated
	phases. However, 
	at particle coalescence	the divergent $1/r^3$ dipolar potential and 
	associated pathological wavefunction hinder
	computational analysis. For a dipolar gas constrained to two dimensions 
we overcome these numerical difficulties by proposing a 
  pseudopotential that is explicitly smooth
	at particle coalescence, resulting in a $2000$-times speedup in
	diffusion Monte Carlo calculations. The
	pseudopotential	delivers the scattering phase
  shifts of the dipolar interaction with an accuracy of $10^{-5}$ and predicts 
the energy of a dipolar gas to
  an accuracy of $10^{-4}\Ef$ in a diffusion Monte Carlo
	calculation.
\end{abstract}

\maketitle

\section{Introduction}

Ultracold atomic gases are an ideal testing ground for many-body quantum 
physics.
Experiments now allow the condensation of particles that carry
either an electric or magnetic dipole moment, and so interact through the 
long-ranged
dipolar interaction in a highly controlled environment 
\cite{Wu12,Werner05,Lahaye07,Neyenhuis12,Park15,
Wang10,Takekoshi14,Ni08,Bismut12,Ni10}. These systems present an ideal 
opportunity to study emergent
strongly correlated phenomena driven by long-range interactions 
\cite{Aikawa14,Cinti10,Parish12,Astrakharchik07,Mazzanti09,Macia11,Macia12,
Macia14,Macia14a,Matveeva13,Matveeva14}.
However, numerical studies of the dipolar interaction are
complicated by the pathological behavior of the wavefunction at
particle coalescence. We propose a pseudopotential for the
dipolar interaction that delivers almost identical scattering properties to the 
original
dipolar interaction, but has a  smooth profile that accelerates diffusion Monte 
Carlo calculations by a factor of $\sim 2000$.

In recent years there have been rapid developments in forming, trapping,
and cooling ultracold atoms and molecules with dipole moments. These 
experiments have involved fermionic \cite{Wu12} or bosonic \cite{Werner05} 
particles with magnetic \cite{Lahaye07} or electric \cite{Neyenhuis12} dipole 
moments, in the continuum \cite{Aikawa14} or a lattice potential 
\cite{Hazzard14}.  For the sake of concreteness we consider a gas of fermionic 
dipolar particles \cite{Wu12,Park15,Wang10}. A particularly appealing geometry 
is
a single component gas of fermions trapped in two dimensions \cite{Ni10}. 
This configuration can suppress the chemical reaction rate of the molecules,
thereby giving sufficient time to relax and study strongly correlated phases 
\cite{Miranda11}, and a strong external field can align the dipoles at an angle 
$\theta$ to the normal to the plane, which allows fine control over the 
interactions between the particles.
The dipolar interaction between the particles is then
$V(r,\phi)=d^2[1-\frac{3}{2}\sin^2\theta(1+\cos2\phi)]/r^3$ where $\phi$ is the 
polar angle in the plane, measured from the projection of the electric field 
onto the plane, $r$ is the inter-particle distance, and $d$ is
the dipole moment.  We focus on the fully repulsive regime of the potential, 
with $\theta\le\theta_\mathrm{c}=\mathrm{arcsin}(1/\sqrt{3})$, where there are 
no bound states.  In the special case $\theta=0$ the potential $V(r,\phi)$ 
reduces to the isotropic form $V(r)=d^2/r^3$.

Theoretical studies of the dipolar gas have provided a rich variety of 
surprises and insights.
Remarkably, even at mean-field level the 
non-tilted ($\theta=0$) system with an isotropic potential is predicted to 
display
an inhomogeneous stripe phase \cite{Sun10,Yamaguchi10} that is robust to the 
inclusion of 
perturbative quantum fluctuations \cite{Parish12}. To extend beyond
the perturbative regime theorists have turned to diffusion Monte Carlo 
\cite{Matveeva12}: however, the divergent dipolar potential and associated 
pathological wavefunction make these simulations difficult to carry out, and 
they have not uncovered evidence of the exotic inhomogeneous stripe phase.

The disagreement between analytical and numerical studies motivates us to 
focus our efforts on improving the modeling of the troublesome dipolar 
potential. Similar difficulties with divergent potentials arise in the
study of the contact and Coulomb interactions, where it has been shown that 
pseudopotentials can
accurately mimic the real interaction \cite{Bugnion14,LloydWilliams15}.
We follow the same prescription to now construct a pseudopotential
that delivers the same scattering physics as the dipolar interaction, but which
is smooth at particle coalescence and so avoids the numerical difficulties 
arising from
pathological behavior near particle coalescence. 

This smoothness will provide benefits in a variety of numerical techniques, 
including configuration interaction methods \cite{Sherrill99}, coupled cluster 
theory \cite{Bartlett07}, and diffusion Monte Carlo (DMC) \cite{Foulkes01}.  
Here we analyze the performance of the pseudopotential by carrying out DMC 
calculations on the dipolar gas to find the ground state energy of the system.  
We find that the proposed pseudopotential delivers ground state energies with 
an accuracy of order $10^{-4}\Ef$, whilst also offering a speedup by a factor 
of $\sim2000$ relative to using the dipolar potential.

We start by studying the two-body scattering problem. In \secref{KatoCusp} 
we analytically solve the wavefunction of the non-tilted $\theta=0$ system near 
to particle coalescence, which offers insights into the numerical difficulties. 
Building on the analytical solution,
in \secref{Derivation} we numerically
solve the two-body problem of scattering from the dipolar potential out to 
larger radii. This provides the
scattering phase shift that we use to calibrate
the scattering from the pseudopotential. Having proposed the pseudopotential,
in \secref{HarmonicTrap}
we test it on a second two-body system: two particles in a parabolic trap.
In \secref{sec:FermiGas} we then demonstrate the use of the pseudopotential to 
study the ground state energy
of the many-body fermionic gas, confirming both the accuracy of the 
pseudopotential and
the computational speedup. In \secref{sec:Tilted} we repeat
the procedure with tilted dipoles, and in \secref{Discussion} discuss future 
applications of the pseudopotential.

\section{Kato-like cusp conditions}\label{KatoCusp}

To develop a pseudopotential for the dipolar interaction we need to
properly understand scattering from the original dipole.  
Working with non-tilted dipoles, we focus on the small radius limit where we 
can solve for the
wavefunction analytically.
This will allow us to demonstrate the pathological behavior of the wavefunction 
and
resultant numerical difficulties, and provide boundary conditions for the full 
numerical
solution of the scattering properties.  Moreover we will calculate
a Kato-like cusp condition, a scheme
to partially alleviate these numerical difficulties for the true dipolar 
potential.

To study the small radius behavior we focus on the two-body problem: two 
identical same-spin
fermions of mass $m$ in their center-of-mass frame with energy $E\ge0$. The 
Hamiltonian in atomic units ($\hbar=m=1$) is
\begin{align}
 \hat{H}\psi(r,\phi)= 
-\nabla^2\psi(r,\phi)+V(\hat{r})\psi(r,\phi)=E\psi(r,\phi),
 \label{SchrodingerEquation}
\end{align}
where $V(r)=d^2/r^3$ is the isotropic dipolar interaction for particle 
separation $r$ and dipole
strength $d$, with characteristic length scale \mbox{$r_0=d^2$}.

\begin{figure}
 \includegraphics[width=\linewidth]{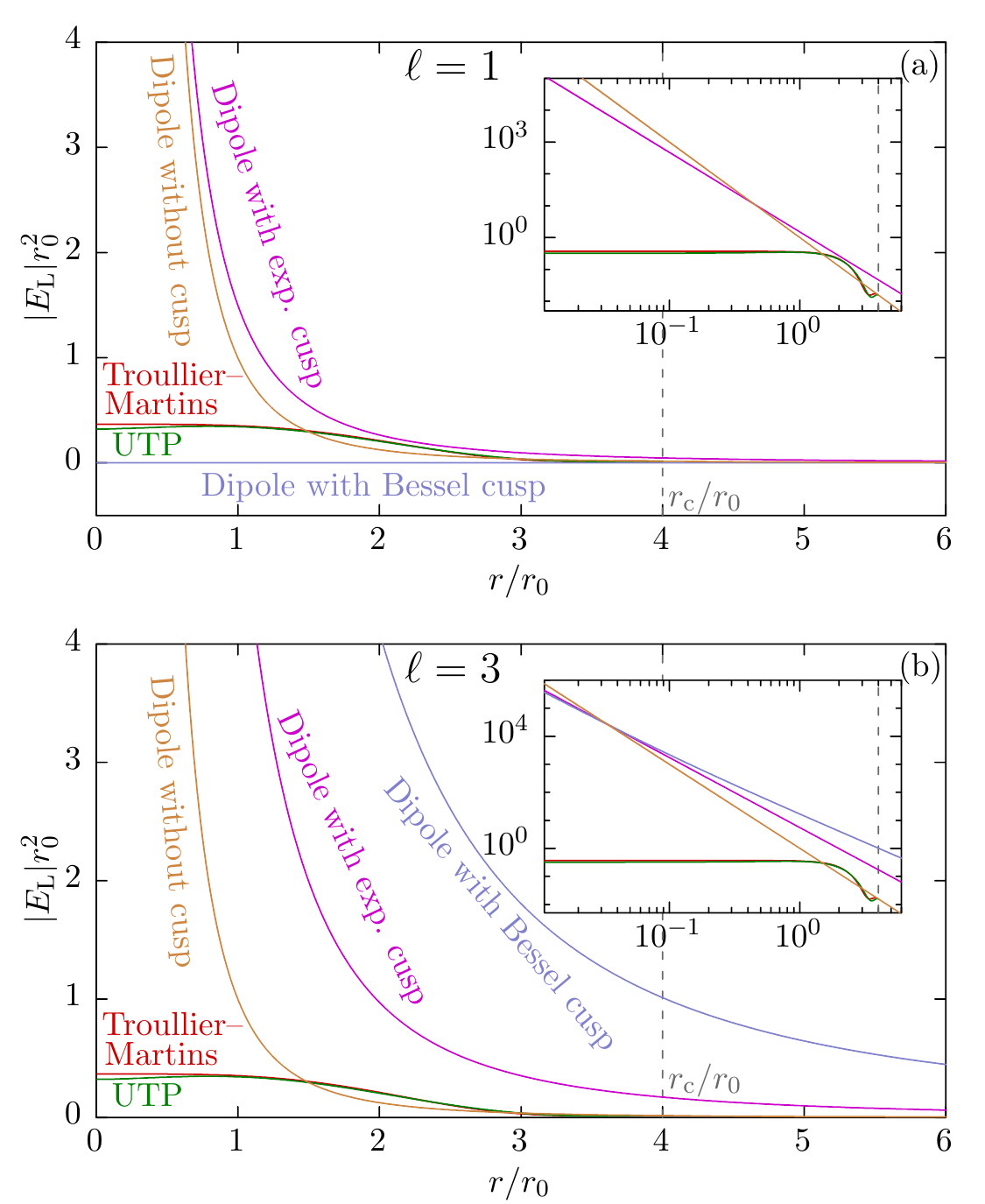}
 \caption{(Color online) (a) The local energy 
$E_\mathrm{L}=\psi^{-1}\hat{H}\psi$ 
as a function of radius in the $\ell=1$ angular momentum channel,
 showing in orange the divergence as $r^{-3}$ when the
dipolar potential is used with the non-interacting wavefunction 
$\psi_{\mathrm{non-int},\ell=1}$.  Also shown in magenta is the local energy 
divergence as 
$r^{-5/2}$ when the dipolar potential is used with a wavefunction with an 
exponential cusp correction $\psi_{\mathrm{exp},\ell=1}$, and in blue the 
exact solution in
this channel, given by a Bessel function cusp correction $\psi_{K_2,\ell=1}$.
In red and green are the local 
energies of Troullier--Martins and ultratransferable (UTP) pseudopotentials, 
respectively, with the non-interacting wavefunction,
which outside of the radius $\rc$ shown by a dashed gray line
join smoothly onto the real dipolar potential. The inset shows the same curves 
on a logarithmic scale.  (b) The local energy in the $\ell=3$ channel, 
demonstrating that the Bessel function cusp correction 
$\psi_{K_2,\ell=3}$ is not accurate in other channels.}
 \label{local_energy}
\end{figure}

A key quantity for Monte Carlo methods is the local energy, 
$E_\mathrm{L}=\psi^{-1}\hat{H}\psi$ \cite{Needs10}. For an eigenstate the 
local energy is constant, and equal to the eigenenergy, whilst for other 
wavefunctions the local energy varies in space.  The foundation of the 
many-body trial wavefunction in our Monte Carlo calculations is a 
non-interacting wavefunction
given by a Slater determinant of plane wave states.  As two particles approach
coalescence their contribution to the wavefunction in each angular momentum 
channel $\ell$ is 
\mbox{$\psi_{\mathrm{non-int},\ell}(r,\phi)=r^\ell \cos(\ell \phi)$},
which is an eigenstate of the two-body non-interacting system.  The Slater
determinant gives such a contribution in every odd angular momentum channel. In
\figref{local_energy} we demonstrate that when this wavefunction is used with 
the dipolar potential the local energy
diverges as $r^{-3}$ in every angular momentum channel. This divergence is
unwelcome as it will make the local 
energy difficult to sample in Monte Carlo calculations, and the variance of the 
samples will give rise to a large statistical uncertainty in the calculated 
energy.

To try to remedy this divergence in the local energy we examine the exact
eigenstates of the two-body Hamiltonian given by \eqnref{SchrodingerEquation},
and then apply our findings to the many-body system.
In the small separation limit where the potential $V(r)$ diverges the 
eigenstates of the Hamiltonian are
\begin{align*}
\psi_\ell(r,\phi)=K_{2\ell}(2\sqrt{r_0/r})\cos(\ell\phi),
\end{align*}
where $K_n(x)$ is a modified Bessel function of the second kind and the
quantum number $\ell$ denotes angular momentum projected onto the polar axis. 
In order to turn the $\ell=1$ part of the non-interacting wavefunction given by
the Slater determinant into
an eigenstate of the Hamiltonian with the dipolar interaction we may multiply 
the Slater determinant by a factor 
$K_{2}(2\sqrt{r_0/r})/r$, which we refer to as a Bessel function cusp
correction.  This gives a wavefunction that is a zero-energy eigenstate 
of the Hamiltonian in the $\ell=1$ channel, as shown in
\figref{local_energy}(a). Similar Bessel function cusp corrections have been 
used previously to study both fermionic and bosonic systems 
\cite{Matveeva12,Astrakharchik07,Macia14a}.

In Monte Carlo calculations we have to pre-multiply the entire Slater
determinant, and so all angular momentum channels present in it, by a single
cusp correction term, and
it is not practical to adapt the cusp correction on the fly to
the relative angular momentum of interacting particles.  However, the Bessel
function cusp correction applied to the two-body wavefunction,
\begin{align*}
\psi_{K_2,\ell}(r,\phi)=r^\ell \cos(\ell \phi) K_2(2 \sqrt{r_0/r})/r,
\end{align*}
is not an eigenstate in any angular momentum channel except $\ell=1$. In other
channels it gives a local energy that diverges as $r^{-5/2}$ in the
$r\to 0$ limit, as shown in \figref{local_energy}(b) for the $\ell=3$ channel.

The improvement of the divergence in the local energy from $r^{-3}$ to 
$r^{-5/2}$ is, in fact, due to the leading-order behavior of the Bessel 
function cusp correction, which goes as $\exp(-2\sqrt{r_0/r})$, independent of
angular momentum.  Accepting that 
we will always be left with an $r^{-5/2}$ divergence of the 
local energy in many-body calculations, we may then just take this leading 
order term to give an
exponential cusp correction, leading to a wavefunction
\begin{align*}
\psi_{\mathrm{exp},\ell}(r,\phi)=r^\ell \cos(\ell \phi) \exp(-2\sqrt{r_0/r}).
\end{align*}
The $r^{-5/2}$ divergence of the local energy with this wavefunction is shown 
in \figref{local_energy} for angular momentum channels $\ell=1$ and $\ell=3$.

The approach of inserting a small radius analytical solution into the many-body
trial wavefunction is well established in
electronic-structure calculations where the small radius behavior of the 
wavefunction
around the $1/r$ divergence in the Coulomb potential
is fixed with the Kato cusp conditions \cite{Kato57,Pack66}.
Following this prescription we can premultiply a many-body 
non-interacting trial wavefunction
by the exponential cusp correction $\prod_{i > j}\exp(-2\sqrt{r_{0}/r_{ij}})$
or Bessel function cusp correction 
$\prod_{i > j}K_{2\ell}(2\sqrt{r_0/r_{ij}})/r_{ij}^\ell$,
where the product is over
all dipoles labeled by $i,j$ and $r_{ij}$ is the dipole-dipole separation.  
Similarly to the two-body case both corrections leave an $r^{-5/2}$ divergence 
in the local energy, which will manifest itself as a major contribution to the 
uncertainty
in the final prediction of the energy. We will revisit the question of cusp 
corrections in a many-body system in
\figref{cutoff_plots}(b), where we show that the simple exponential
cusp correction gives similar values for the variance in the local energy
to a full Bessel function cusp correction.

In order to study the interacting-dipole system further we turn to the 
construction of
pseudopotentials \cite{Bugnion14,Troullier91} that capture the physics of
the system whilst delivering the smooth and non-divergent local energy values 
shown in \figref{local_energy}.

\section{Derivation of the pseudopotentials}\label{Derivation}

To construct a pseudopotential for the dipolar interaction we continue with the 
two-body scattering problem
of two indistinguishable fermions in their center-of-mass frame, studying the
Schr\"odinger \eqnref{SchrodingerEquation}. We seek a pseudopotential that
is smooth and non-divergent to accelerate numerical
calculations.  We also require it to reproduce the correct two-body scattering 
physics over the range of scattering energies present in a Fermi gas with Fermi 
energy $\Ef$, which guarantees that the pseudopotential will properly capture 
two-body effects in the system.  As we will be considering two-body processes 
we again work in the center-of-mass frame, with the Hamiltonian given by 
\eqnref{SchrodingerEquation}.

We first turn to the
Troullier--Martins~\cite{Troullier91} formalism that has been widely used and
rigorously tested in the literature to
construct attractive electron-ion
pseudopotentials~\cite{Trail05,Trail05a,Heine70,Hamann79,Zunger79,Bachelet82},
but which may be adapted \cite{Bugnion14} to the current problem of two 
identical fermions as
detailed in \appref{TroullierMartins}.  This method
creates a pseudopotential with the exact dipolar potential outside of a
cutoff radius $\rc$ and a polynomial potential within it, constructed to be
smooth up to second derivative at $\rc$.  The Troullier--Martins method
guarantees that the scattering properties of the pseudopotential will be
exact at one particular calibration energy $\Ec$.  We choose the calibration 
energy to be the average
scattering energy of two fermions in a non-interacting Fermi gas. In
\appref{CalibrationEnergy} we show that this calibration energy is $\Ec=\Ef/4$.

For the scattering of two indistinguishable fermions the Pauli principle
guarantees that there will be no $s$-wave contribution to the scattering.  We
therefore construct the Troullier--Martins pseudopotential by focusing on a 
scattering wavefunction in the $p$-wave,
$\ell=1$, channel.  The
functional form of the pseudo-wavefunction in this channel is
\begin{align}
 \psi_{\ell=1}(r,\phi)=
 \begin{cases}
  \exp[p(r)]\, r\cos(\phi)\punc{,}&r<\rc\punc{,}\\
  \psi_{\mathrm{dipole},\ell=1}(r,\phi)\punc{,}&r\ge\rc\punc{,}
 \end{cases}
 \label{PseudoWfn}
\end{align}
where the polynomial $p(r)=\sum_{i=0}^6 c_{i}r^{2i}$, and the wavefunction 
$\psi_{\mathrm{dipole},\ell=1} (r,\phi)$ is
calculated by numerically solving \eqnref{SchrodingerEquation} using the exact
dipolar potential at the calibration energy $\Ec$.  As explained in
\appref{TroullierMartins} the coefficients $c_i$ are
calculated by requiring continuity of the pseudo-wavefunction and its first 
four
derivatives at $\rc$, as well as matching the net density inside $\rc$, and
requiring the pseudopotential to have zero gradient and curvature at the origin.

The choice of $\rc$ is motivated by the physics we wish to study: a longer
cutoff radius allows a smoother potential that gives efficient numerics, but 
being less similar to the real potential has less accurate phase shift errors. 
In many-body systems the longer cutoff radius will also increase the 
probability of having three or more particles within the cutoff radius, which 
the pseudopotential is not designed to be able to accurately model.  For our 
two-body scattering system we take $\kf \rc=2$.

\begin{figure}
\includegraphics[width=\linewidth]{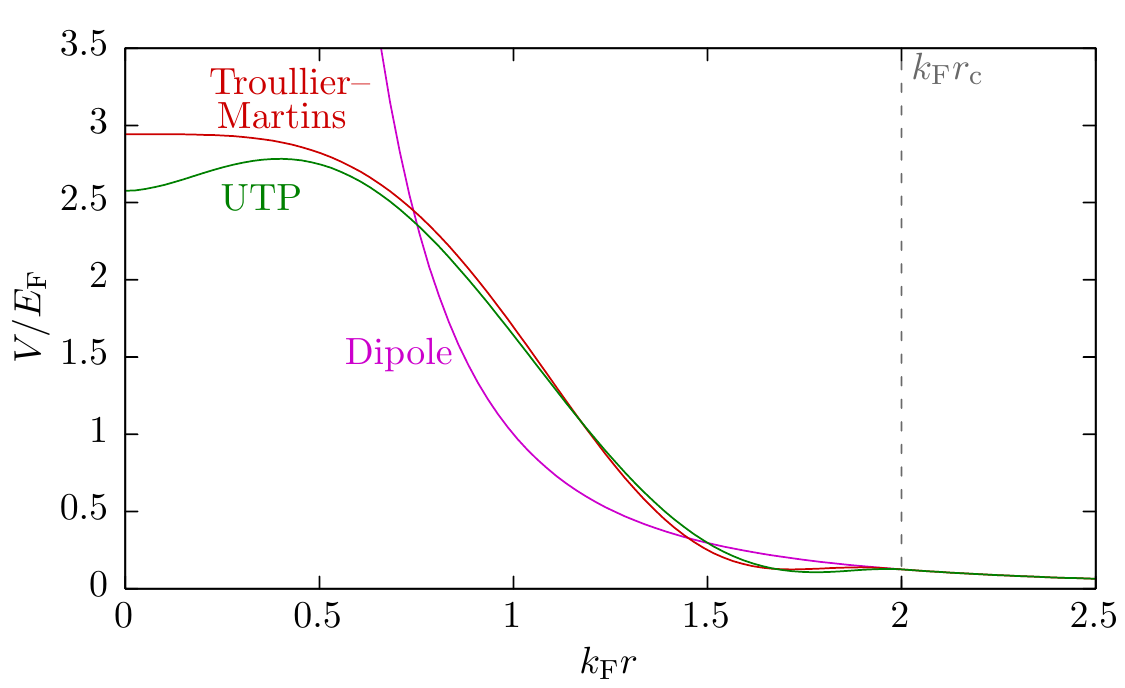}
\caption{(Color online) The dipolar potential, and Troullier--Martins and UTP 
pseudopotentials. The gray vertical line indicates $\rc$, the pseudopotential 
cutoff radius.}
\label{potentials}
\end{figure}

The exponentiated polynomial form of the pseudo-wavefunction in 
\eqnref{PseudoWfn} means that the Schr\"odinger \eqnref{SchrodingerEquation}
may be analytically inverted to give the pseudopotential as
\begin{align}
 V_\mathrm{T\textendash M}(r) = 
  \begin{cases}
   \Ec + \frac{3}{r} p' + p'^2 + p''\punc{,} & r<\rc\punc{,} \\
   d^2/r^3\punc{,} & r\ge\rc,
  \end{cases}
\label{PseudoPot}
\end{align}
where the primes denote differentiation with respect to $r$. This 
pseudopotential is shown in red in
\figref{potentials} for interaction strength $\kf r_0=1/2$. It is non-divergent 
at particle coalescence and smooth where it joins onto the real dipolar 
potential at $r=\rc$.  This pseudopotential gives rise to the local energy 
$E_\mathrm{L}$ shown in \figref{local_energy}.  The smooth and finite local 
energy at $r<\rc$ is a dramatic improvement over the divergent local energy 
from our trial wavefunction with the dipolar potential, and this non-divergence 
should lead to improved statistics and efficiency in many-body simulations.

\begin{figure}
\includegraphics[width=\linewidth]{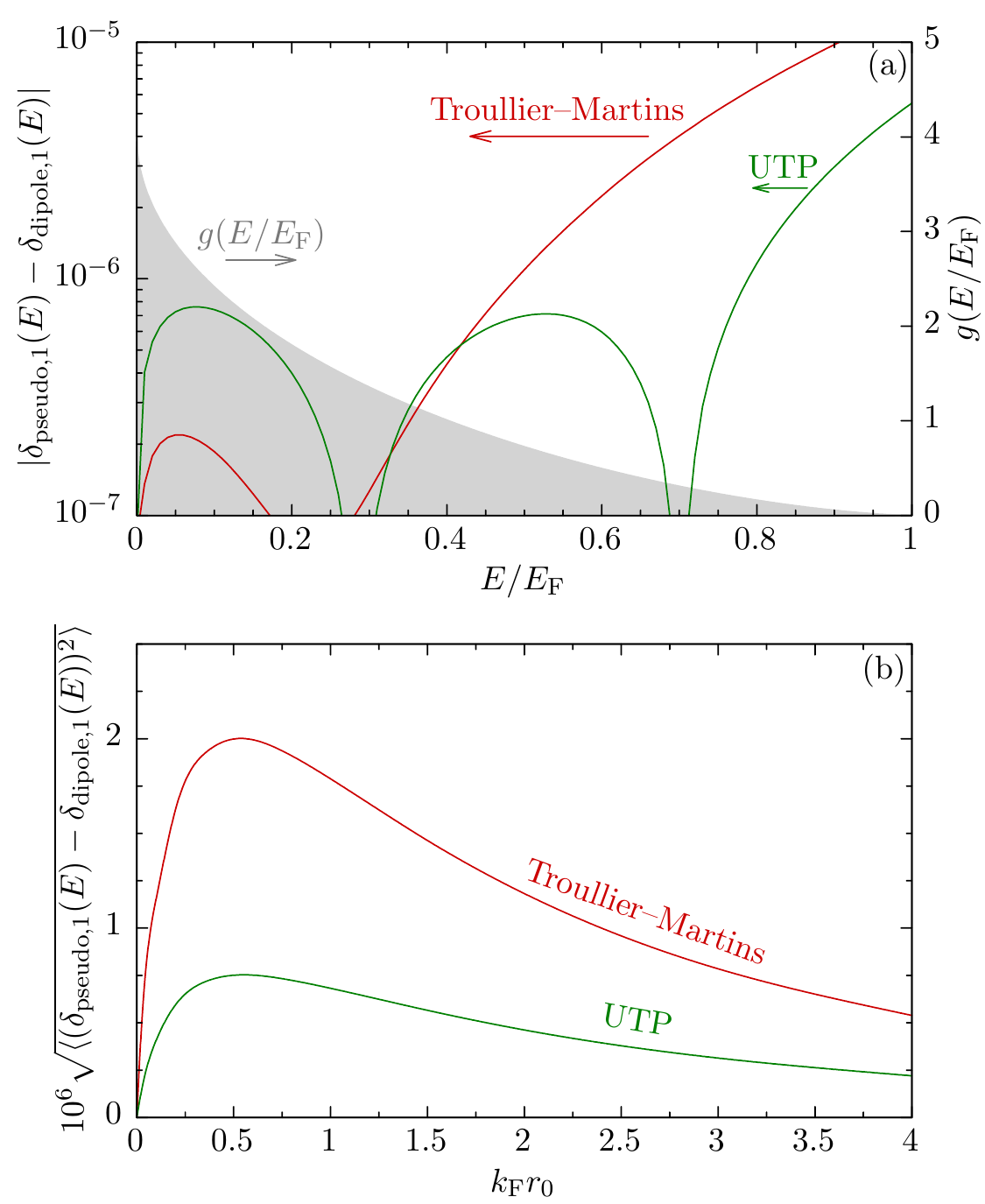}
\caption{(Color online) (a) The error in the scattering phase shift 
$|\delta_{\mathrm{pseudo},1}(E)
  -\delta_{\mathrm{dipole},1}(E)|$. The
  filled gray curve is the density of scattering states $g(E)$ in
  the two-body Fermi sea on a linear scale.  (b) The
  root-mean-squared error in the scattering phase shift as a function of
  interaction strength.}
\label{phase_differences}
\end{figure}

To measure the accuracy of our pseudopotentials we calculate the phase shift in 
the wavefunction
\begin{align}
 \delta_{\psi,\ell}(E)=\frac{1}{2 \pi}\mathrm{arccot}
 \left[\frac{1}{\sqrt{E}}
\left(\frac{\psi_\ell'(\rc,\phi)}{\psi_\ell(\rc,\phi)}+\frac{2\ell+1}{2\rc}
\right)\right]
 \label{phases}
\end{align}
imparted by a two-body scattering process, where $\delta_{\psi,\ell}$ is 
evaluated at the cutoff radius $\rc$ because any difference in phase shift must 
be accumulated in the region $r<\rc$ where the potentials differ.  The 
difference
between the scattering phase shift for the Troullier--Martins pseudopotential 
and the exact phase shift from the dipolar
interaction is shown in red in \figref{phase_differences}(a) as a function of 
scattering energy, evaluated at $\kf r_0=1/2$.  The scattering phase shift of 
the Troullier--Martins pseudopotential is exact at the calibration energy, and 
accurate to order $10^{-5}$ over the range of scattering energies in a Fermi 
sea.  

Although the Troullier--Martins pseudopotential captures the exact scattering
properties at the calibration energy, it deviates at all other energies, with 
the leading order deviation around the calibration energy going as $(E-\Ec)^2$ 
\cite{Bugnion14}.  A
natural extension to the Troullier--Martins formalism is to find a
pseudopotential that minimizes this deviation in the phase shift over all the
possible relative energies of pairs of particles in a Fermi gas.  We
derive such a pseudopotential here, referring to it as an
``ultratransferable pseudopotential'' (UTP).

The UTP \cite{Bugnion14} is identical to the dipolar potential outside a cutoff 
radius $\rc$,
but has a polynomial form inside the cutoff,
\begin{align*}
 V_\mathrm{UTP}(r)\!=\!\frac{d^2}{\rc^3}\!\!
 \begin{cases}
  \begin{array}{l}
   1+3\left(1-\frac{r}{\rc}\right)\left(\frac{r}{\rc}\right)^2+\neweqnline
	\!\!\! \left(1\!-\!\frac{r}{\rc}\right)^{\!2}		
  \!\! 
\left[v_1\!\!\left(\frac{1}{2}\!+\!\!\frac{r}{\rc}\right)\!\!+\!\!\displaystyle
\sum_{i=2}^{N_v}v_i \!\!\left(\frac{r}{\rc}\right)^{\!\!i}\right]\!\!\!\punc{,}
  \end{array}
  &\!\!\!\!\!\!\! \begin{array}{c}\vspace{4.5pt}\neweqnline 
r<\rc\punc{,}\end{array} 
  \vspace{8pt}\\
\rc^3/r^3\punc{,}&\!\!\!\!\!\! r\ge\rc\!\punc{,}
 \end{cases}
\end{align*}
with $N_v=3$.  The term $1+3(1-r/\rc)(r/\rc)^2$ guarantees that the potential 
and its first derivative are continuous at $r=\rc$.  In the next term, the 
expression $(1-r/\rc)^2$ also ensures continuity of the potential at the cutoff 
radius, and $v_1(1/2+r/\rc)$ constrains the potential to have zero derivative 
at the origin.  This ensures that the pseudo-wavefunction is smooth, easing the 
application of numerical methods.

To determine the coefficients $\{v_i\}$ we minimize the total squared
error in the phase shift over all the possible pairs of interacting particles 
in a
Fermi gas 
\begin{align}
&\left\langle\left|\delta_{\mathrm{UTP},\ell}\left(E\right)-\delta_{\mathrm{dipo
le},\ell}\left(E\right)\right|^{2}\right\rangle\neweqnline
&=\int \left|\delta_{\mathrm{UTP},\ell}\left( 
E\right)-\delta_{\mathrm{dipole},\ell}\left( E\right)\right|^{2} g(E/\Ef)\, \dd 
E/\Ef\punc{,}
\label{UTPfinder}
\end{align}
where 
\begin{align*}
g(x)=4 - \frac{8}{\pi}\left( \sqrt{x(1-x)} + \mathrm{arcsin} \sqrt{x} \right) 
\end{align*}
is the density of scattering states in energy (see \appref{CalibrationEnergy} 
and Reference~\cite{Lu12}), shown in \figref{phase_differences}(a). The density of 
scattering states decreases as a function of energy due to the finite size of 
the Fermi sea of scattering particles limiting the available range of 
scattering energies.  We primarily work in the leading-order $\ell=1$ angular 
momentum channel.
The UTP formalism is capable of creating pseudopotentials that are accurate
in several angular momentum channels by summing over them in
Equation~\eqref{UTPfinder} whilst accounting for the occupation of the channels 
that goes as
$1/\sqrt{(2\ell+1)!!}$ \cite{LloydWilliams15}, which strongly suppresses the
effect of all the channels above $\ell=1$. The total squared phase shift error
\eqnref{UTPfinder} is numerically minimized with respect to the $v_i$ to create 
our UTP.

The scattering phase shift behavior of the UTP is shown in
\figref{phase_differences}(a).  Although it is less accurate than the 
Troullier--Martins pseudopotential at
the Troullier--Martins calibration energy, the UTP is more accurate at higher 
incident energies. At zero scattering energy both pseudopotentials are exact, 
as the scattering particles never penetrate the region $r<\rc$ where the 
pseudopotentials deviate from the real dipolar interaction.

In \figref{phase_differences}(b) we show the average phase shift error in the 
pseudopotentials as a function of interaction strength.  At its worst the 
Troullier--Martins pseudopotential has an average accuracy of $2\times 
10^{-6}$, whilst the average UTP accuracy is always better than $1\times 
10^{-6}$. Over a broad range of interaction strengths the UTP is more accurate 
than the Troullier--Martins pseudopotential, but both are exact at $\kf r_0=0$ 
where the particles do not interact. At high interaction strengths the 
pseudopotentials become highly accurate, as the increasing interaction strength 
effectively rescales the potential size, and so for a given range of scattering 
energies the particles will be kept further apart and so less strongly probe 
the region $r<\rc$ where the potentials differ.  We also note that a further 
advantage of the UTP is that at high interaction strengths, $\kf r_0>4$ with 
$\kf \rc=2$, it is not possible to solve the system of equations defining the 
Troullier--Martins pseudopotential, whilst it is still possible to derive a UTP.

Having constructed two different pseudopotentials and demonstrated their 
accuracy in a homogeneous two-body setting, we now test their flexibility by 
solving an inhomogeneous two-body system.

\section{Two fermions in an harmonic trap}\label{HarmonicTrap}

We have developed pseudopotentials that exhibit the correct scattering
properties for an isolated two-body system.  To test them we turn to the
experimentally realizable \cite{Murmann15,Moses15} configuration of two
fermionic dipolar particles aligned by an external field and held in a
circularly symmetric two-dimensional harmonic well with trapping frequency 
$\omega$.  Given that the identical fermions must be in
different single-particle states of the harmonic trap the non-interacting 
energy of the reduced system is $2\omega$.  This system is a good place to test 
our pseudopotentials as it has a non-trivial background potential, but at the 
same time is still simple enough to solve accurately with the real dipolar 
potential.

\begin{figure}
\includegraphics[width=\linewidth]{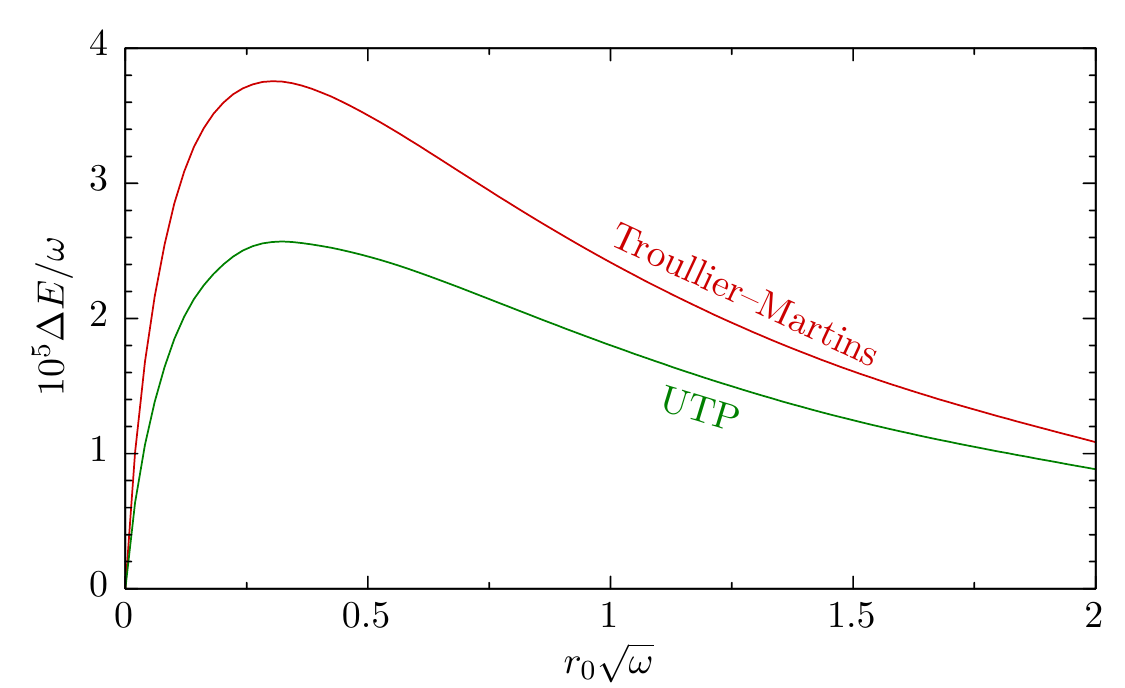}
\caption{(Color online) The deviation of the energy of two particles in an 
harmonic trap as
  calculated using both Troullier--Martins and UTP pseudopotentials from that 
calculated using the exact
  dipolar potential, as a function of interaction strength.}
\label{harmonic_trap}
\end{figure}

We calculate the energy of two particles held in such a trap by solving the 
Schr\"odinger equation for the relative motion in the system,
\begin{align}
-\nabla^2\psi+\frac{1}{4}\omega^2r^2\psi+V(r)\psi=E\psi\punc{,}
\label{SchrodingerEquationHarmonic}
\end{align}
with $V(r)$ set as either the exact dipolar potential or a pseudopotential.  We
solve the system in the lowest-energy $\ell=1$ angular momentum channel 
available to identical fermions, calibrating the Troullier--Martins 
pseudopotential at $\Ec=(2\omega)/4=\omega/2$ by analogy to the homogeneous 
system.  For the cutoff radius $\rc$ we choose the
characteristic width of the trap, $1/\sqrt{\omega}$.

The energy differences between the pseudopotential and exact dipolar
solutions to Equation~\eqref{SchrodingerEquationHarmonic} are shown in
\figref{harmonic_trap} as a function of interaction strength.  Approaching zero 
interaction strength the form of the interaction potential has diminishing 
impact, and so the difference in energies goes to zero; and in the 
high-interaction strength limit the particles are kept further apart by the 
strong potential, so less strongly probe $r<\rc$ where the potentials differ 
and again the error in the ground state energy becomes negligible.  At 
intermediate interaction strengths $r_0 \sqrt{\omega}\approx1/4$ the 
pseudopotentials are still accurate to order $10^{-5}\omega$, which exceeds the 
$\sim10^{-4}\omega$ accuracy attainable in exact diagonalization 
\cite{Bugnion13} and many-body
quantum Monte Carlo calculations \cite{Matveeva12,Bugnion14,Conduit09}. The UTP 
provides an improvement in accuracy over the Troullier--Martins pseudopotential 
at all interaction strengths.

\section{Fermi gas}\label{sec:FermiGas}

Having demonstrated that the Troullier--Martins and UTP pseudopotentials are
accurate tools for studying both scattering and inhomogeneous trapped two-body 
systems, we are
well placed to test the pseudopotentials in a many-body system: a gas of 
fermionic
dipolar particles. The particles are constrained to lie in two dimensions
with all their dipole moments aligned normal to the plane, which has been 
suggested for experimental investigation \cite{Ni10}.  We use
diffusion Monte Carlo (DMC) calculations to study the system, using the 
{\sc{casino}}
code~\cite{Needs10}.

\subsection{Formalism}
\label{formalism}

Our DMC calculations use 81 particles per simulation cell and
a Slater--Jastrow type wavefunction $\Psi = \expe^J D$. Here $D$ is a Slater
determinant of plane-wave orbitals, with wavevectors given by the reciprocal 
lattice vectors of our simulation cell, and the Jastrow factor $\expe^J$ 
describes the
interparticle correlations \cite{Rios12}, with
\begin{align}
J=\sum_{i\ne j}\Bigg(&\sum_{k=0}^{N_\mathrm{u}}u_{k}r_{ij}^{k}
\left(1-\frac{r_{ij}}{L}\right)^3\Theta\left(L- 
r_{ij}\right)\neweqnline+&\sum_{\vect{G}}p_{|\vect{G}|}\cos(\vect{G}\cdot\vect{r
}_{ij})\Bigg)\punc{,}
\label{Jastrow_factor}
\end{align}
where the first sum runs over all particles labeled $i,j$ with separation 
$\vect{r}_{ij}$, $N_\mathrm{u}=7$, and the $\vect{G}$ vectors are the 36 
shortest reciprocal lattice vectors (first 8 sets of equal-length reciprocal 
lattice vectors).  The cutoff function $(1-r_{ij}/L)^3$ ensures that the 
wavefunction's first two derivatives go smoothly to zero at a radius $L$, 
chosen to be the Wigner-Seitz radius of the simulation cell. Calculations with 
the exact
dipolar interaction have a cusp correction term in the Jastrow factor,
using the exponential form $\prod_{i > j}\exp(-2\sqrt{r_{0}/r_{ij}})$ as 
discussed in \secref{KatoCusp}.  We also
test the Bessel function cusp correction proposed in 
Reference~\cite{Matveeva12}. The coefficients $\{u_k\}$ and 
$\{p_{|\vect{G}|}\}$ are 
optimized in a variational Monte Carlo calculation, and then this optimized 
wavefunction is taken as the trial wavefunction for a DMC calculation to 
evaluate the ground state energy.  

We use 4000 particle configurations in DMC,
and by running tests with 2000, 4000, and 8000 configurations checked 
that 4000 configurations gives results within statistical uncertainty of the 
extrapolated result with an infinite number of configurations.  Similarly we 
checked that our system of $81$ particles gave similar results to systems of 
45 and 145 particles, although a full extrapolation of results to the 
thermodynamic limit is not necessary to verify the accuracy of short ranged
pseudopotentials, and so not a focus of this work.
We did however correct the non-interacting energy of the 
system to the result of the infinite system, to reduce finite-size effects in 
the calculation \cite{Bugnion14,Pilati10}.

To evaluate the dipolar interaction we explicitly sum over pairs of particles 
within a distance $\bigRs$ of each other, and then include the effect of 
particles further apart by integrating over them, assuming a uniform particle 
density.  By taking $\bigRs$ as $\sim 18$ simulation cell lattice vectors the 
error due to the finite value of $\bigRs$ is smaller than $10^{-6}\Ef$, and 
therefore negligible compared to our DMC statistical errors 
\cite{Matveeva12,Drummond11}. 

\begin{figure}
\includegraphics[width=\linewidth]{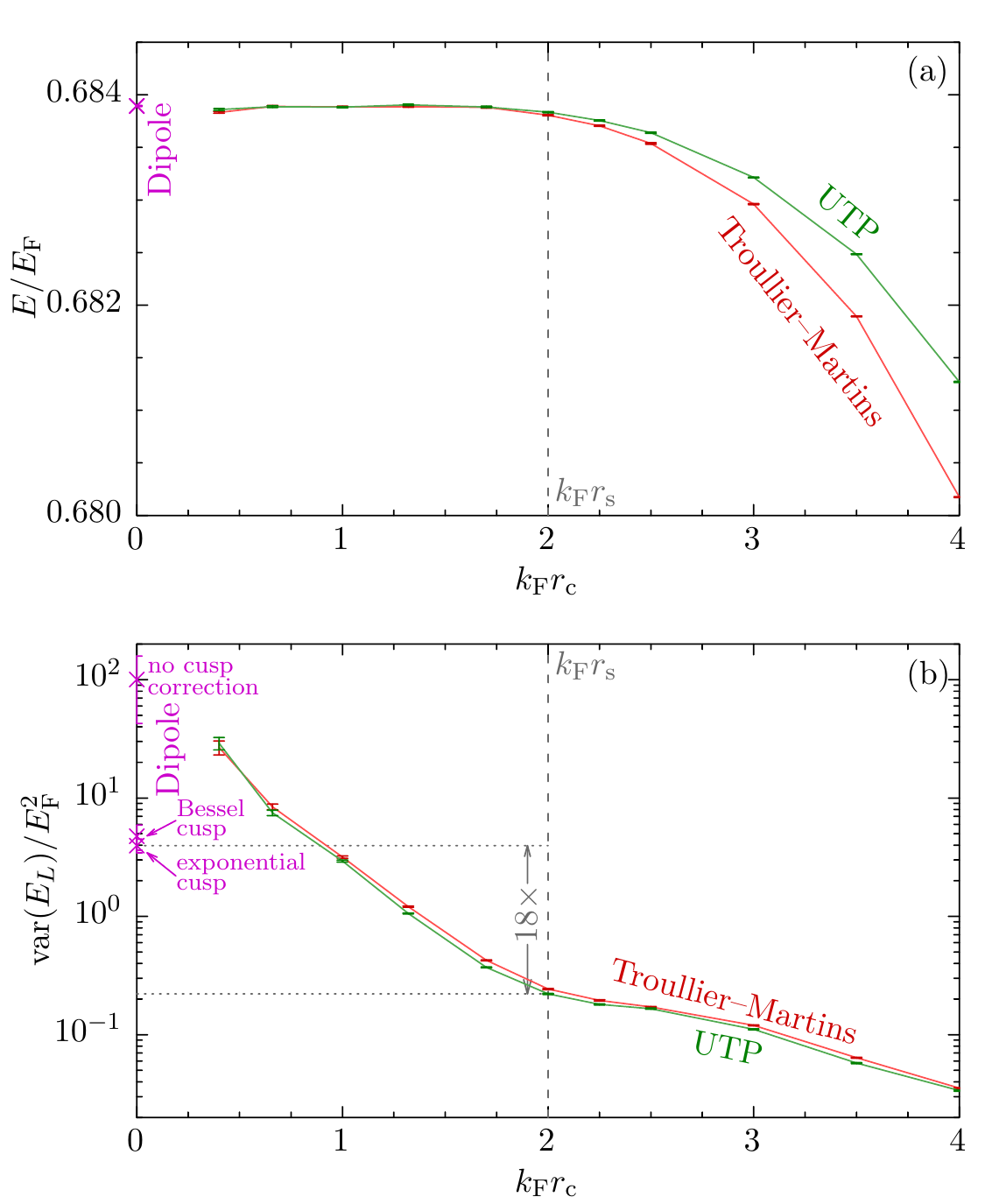}
\caption{(Color online) (a) The variation of the energy per particle in the 
Fermi gas with
  pseudopotential cutoff radius, calculated using DMC. The red points are for 
the Troullier--Martins
  pseudopotential, the green a UTP pseudopotential, and the magenta point is 
the exact dipolar potential.  Stochastic error bars are of order $10^{-5}\Ef$.
	The vertical dashed line denotes the recommended cutoff radius.
  (b) The variance in the individual local energy samples (as seen in 
\figref{local_energy}) taken during a DMC calculation
using the pseudopotentials.  Also shown are results for the dipolar potential 
both with and without Kato-like cusp corrections applied.}
\label{cutoff_plots}
\end{figure}

In order to analyze the accuracy of our pseudopotentials in capturing the 
dipolar gas, we start by fixing the interaction strength and investigate the 
dependence of the accuracy on the cutoff radius $\rc$.  Having selected a 
cutoff radius we then study the effect of the DMC timestep $\tau$, and finally 
present results at a variety of interaction strengths.

In simulations using the pseudopotentials decreasing the cutoff radius 
makes the calculation more accurate by increasing the similarity to the real 
potential and reducing the likelihood of three-body interactions within the 
cutoff radius.  This is shown in
\figref{cutoff_plots}(a), calculated at $\kf
  r_0=1/2$ with timestep $\tau \Ef=0.0092$. However, this increased similarity 
to the dipolar potential also has the effect of increasing the variance in the 
individual local energy samples taken during the simulation, as shown in
\figref{cutoff_plots}(b), which the runtime of a DMC calculation is 
proportional to \cite{Foulkes01}. When using the pseudopotentials a balance 
therefore has to be struck between accuracy and speedup: we choose to take the 
cutoff radius
as equal to $r_\mathrm{s}$, the density parameter that corresponds to the 
average
separation of particles.  This gives DMC calculations with an accuracy of order 
$10^{-4}\Ef$, whilst as shown in \figref{cutoff_plots}(a) this accuracy quickly 
drops off for $\rc >r_\mathrm{s}$. 

In \figref{cutoff_plots}(b) we compare the variance in the individual local 
energy samples from the 
pseudopotentials to that from the real dipolar potential, using wavefunctions
both with and without Kato-like cusp corrections applied.  The two forms of
cusp correction, the Bessel function cusp correction proposed for this system
in Reference~\cite{Matveeva12} and our simpler exponential cusp correction, 
agree to within statistical uncertianty.  As discussed in \secref{KatoCusp} 
this is because both give rise to $r^{-5/2}$ divergences in the local
energy, which are preferable to the higher variance in the local energy from
the bare dipolar potential, which diverges as $r^{-3}$. The source of this 
divergence is, however, more transparent for the exponential cusp correction
than the Bessel function cusp correction,
and so we use the exponential form in the rest of our calculations.

Taking $\rc = r_\mathrm{s}$ for the cutoff radius 
gives an 18-times reduction in the variance of the local energy samples of the 
many-body system using a pseudopotential when compared to using the real 
dipolar interaction with a
Kato-like exponential cusp correction. To get the same statistical error in our
results we 
therefore need to take 18 times fewer samples, leading to an 18-times 
statistical speedup in calculations.

\begin{figure}
\includegraphics[width=\linewidth]{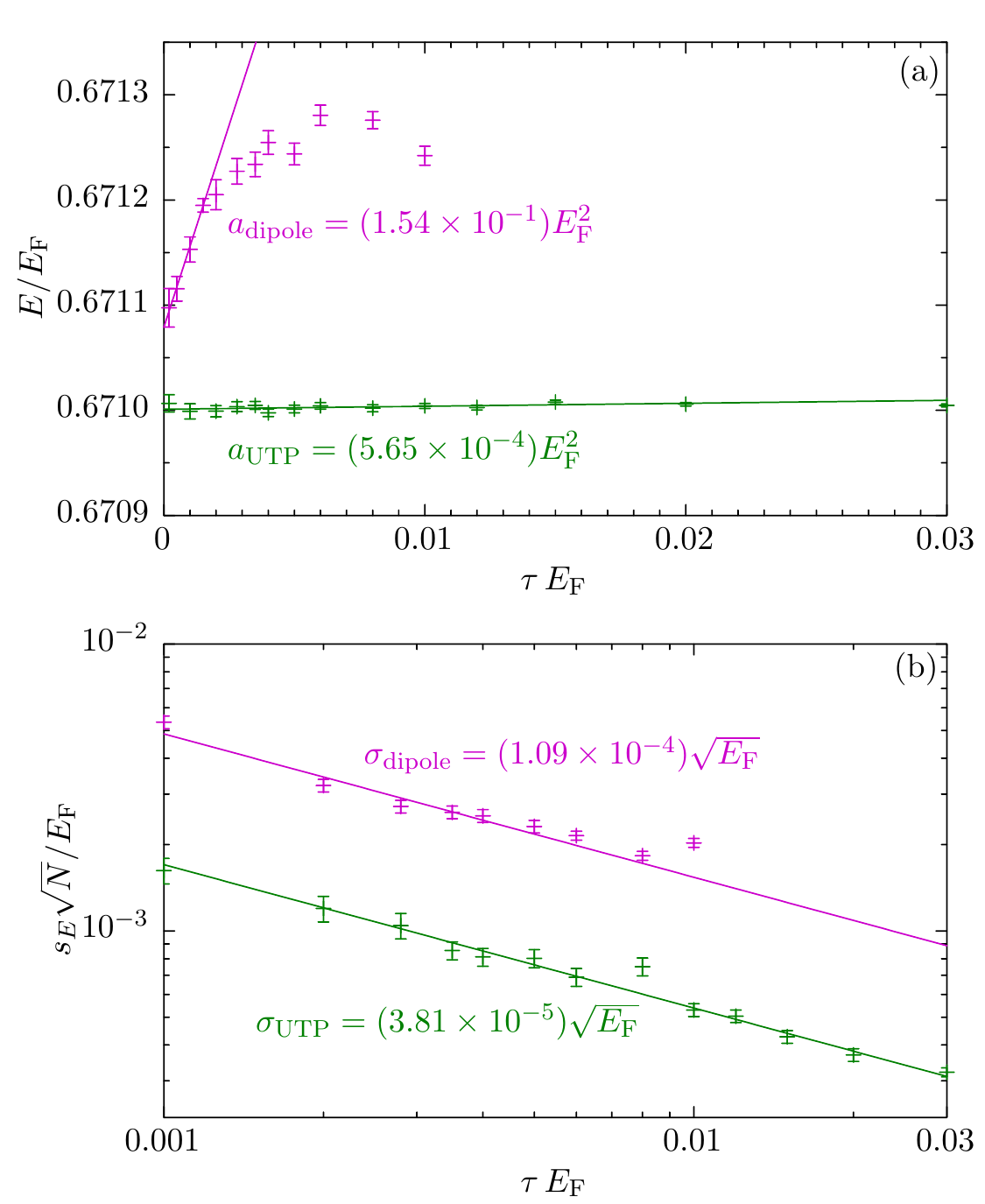}
\caption{(Color online) (a) The variation of the energy per particle in the 
Fermi gas with timestep $\tau$.  The magenta points are using the exact dipolar 
potential, and the green points using a UTP pseudopotential.  The error bars 
show DMC stochastic errors, and are of order $10^{-5}\Ef$.  Fitted values of 
the linear error parameters $a$ (see main text) are also given. (b) The 
standard error $s_E$ in the energy per particle in the Fermi gas, for both the 
dipolar potential and UTP pseudopotential.  Values of the fitting parameters 
$\sigma$ for a $1/\sqrt{\tau}$ fit are also given for each.}
\label{dtdmc}
\end{figure}

There is however an additional speedup benefit from using the pseudopotential. 
The random walk in the DMC calculations is performed at a finite timestep 
$\tau$ \cite{Needs10,Lee11}.  The use of a short-time approximation in the DMC 
algorithm gives rise to a linear dependence of the final estimate of the energy 
on $\tau$ \cite{Needs10}.  If we were to use a short timestep to remove this 
systematic error the DMC  walkers would not be able to move far in 
configuration space in each step, giving rise to serial correlations in the 
calculated values of the energy, and an explicit $\tau^{-1/2}$ dependence of 
the statistical standard error in the energy \cite{Rothstein88}. These two 
competing effects are shown in \figref{dtdmc}(a) and \figref{dtdmc}(b) 
respectively for our Fermi gas at $\kf r_0=1/2$.  The dependence on the energy 
on $\tau$ is both flatter when using the UTP compared to the dipolar potential, 
and also retains its linear form out to larger timesteps: this is advantageous 
as it allows the use of longer timesteps in DMC, which is more efficient.  
\figref{dtdmc}(b) confirms the $\tau^{-1/2}$ dependence of the standard error 
in the energy, and that the smoothness of the UTP delivers a smaller standard 
error.

We express the linear short-time approximation as giving an offset in the 
calculated energy of $a \tau$, where $a$ is a fitting parameter, and the serial 
correlations as giving a variance in the energy of $s_E^2=\sigma^2 N^{-1} 
\tau^{-1}$, with  $\sigma$ being a fitting parameter.  The statistical error 
can  be reduced by taking more samples $N$ \cite{Foulkes01}. We can then 
express the expected value of the square error in the energy as being 
distributed to leading order as a Gaussian \cite{Trail08,LloydWilliams15}
\begin{align}
\left\langle \Delta E^2 \right \rangle &= \int \Delta E^2 \expe^{-\frac{(\Delta 
E - a \tau)^2}{2 \sigma^2 N^{-1} \tau^{-1}}} \dd (\Delta E)\neweqnline
&=a^2 \tau^2 + \sigma^2 N^{-1} \tau^{-1}.
\label{timestep_error}
\end{align}
The expected square error in the energy is minimized at the optimal timestep
\begin{align*}
\tau_\mathrm{optimum} = \left( \frac{1}{2}\frac{\sigma^2}{a^2} \frac{1}{N} 
\right)^{1/3}\punc{,}
\end{align*}
and substituting this into \eqnref{timestep_error}, the ratio of the number of 
steps required to give the same expected square error in the energy when using 
the dipolar potential and the UTP is
\begin{align}
\frac{a_\mathrm{dipole} \sigma_\mathrm{dipole}^2}{a_\mathrm{UTP} 
\sigma_\mathrm{UTP}^2}.
\end{align}
For the values of the fitting parameters $a$ and $\sigma$ in \figref{dtdmc} 
this gives a ratio of required number of steps and hence speedup when using the 
pseudopotential of $\sim2230$.  This value for the speedup includes the 
variance difference of $18$ that was found with the recommended value of $\rc$, 
the remainder coming from the improvement of the finite timestep behavior when 
using the pseudopotential.  

\begin{figure}
\includegraphics[width=\linewidth]{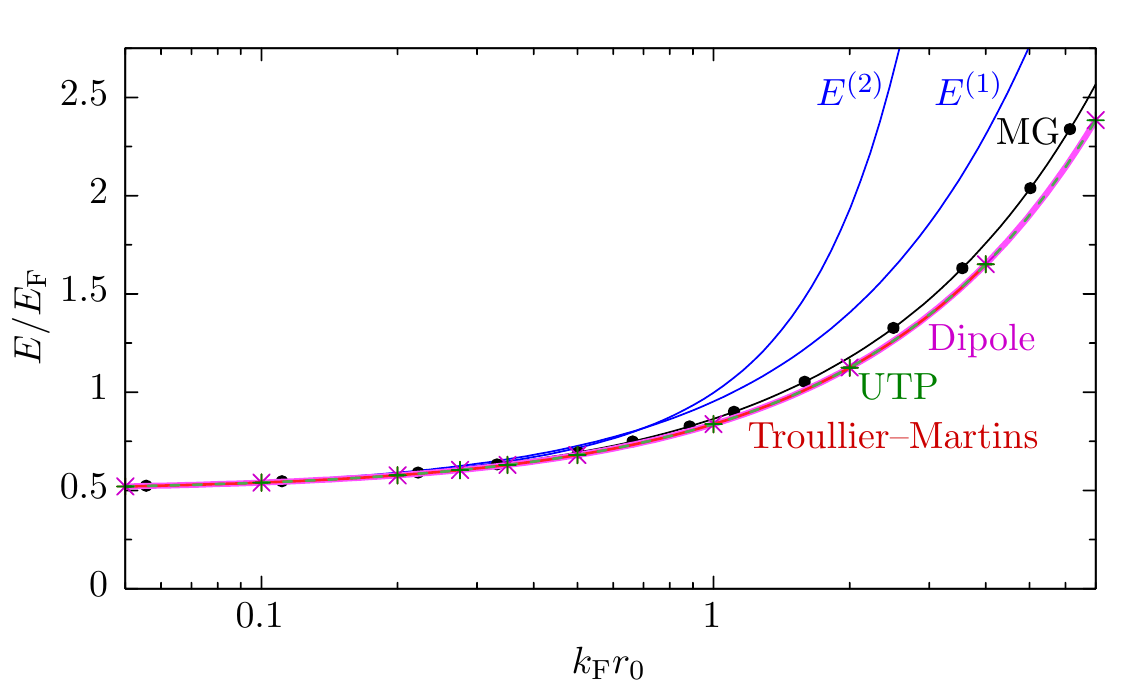}
\caption{(Color online) The equation of state of the 2D isotropic, homogeneous 
dipolar gas.  The blue curves show the first- and second-order perturbation 
theory ($E^{(1)}$ and $E^{(2)}$) equations of state~\cite{Lu12}, and our DMC 
data are shown in: magenta, for the dipolar potential; red for the 
Troullier--Martins pseudopotential calibrated at $\Ef/4$; and green for a UTP.  
The latter three curves overlie each other to within the width of the plotted 
lines.  Stochastic error bars are of order $10^{-5}\Ef$.  The black circles 
show data from DMC calculations using the dipolar potential by 
Matveeva~and~Giorgini (MG) in Reference~\cite{Matveeva12}.}
\label{dmc_energies}
\end{figure}

Use of a second order propagator in the DMC 
algorithm might improve the efficiency of the calculations by allowing the use 
of a longer timestep than was possible here \cite{Mella00,Sarsa02,Chiesa03}. 
In a second 
order DMC algorithm the square error in the energy would take the form 
\mbox{$\left 
\langle \Delta E^2 \right \rangle =b^4 \tau^4 + \sigma^2 N^{-1} \tau^{-1}$}.
The parameter $b$, which is zero if the exact wavefunction is used in DMC, 
should grow with the standard deviation in the local energy. This same effect 
is seen in \figref{dtdmc}(a) and in the results of
Reference~\cite{LloydWilliams15}.  We therefore expect 
$b_\mathrm{UTP}<b_\mathrm{dipole}$, and saw above that 
$\sigma_\mathrm{UTP}<\sigma_\mathrm{dipole}$.  With this form of the square 
error in the 
energy, the speedup when using the pseudopotential relative to the real dipolar
potential would take the form 
\mbox{$b_\mathrm{dipole} \sigma^2_\mathrm{dipole}/(b_\mathrm{UTP} 
\sigma^2_\mathrm{UTP})$}.  We obtain the same statistical speedup as in the 
linear case from the factor $\sigma^2_\mathrm{dipole}/\sigma^2_\mathrm{UTP}$, 
and the ratio $b_\mathrm{dipole}/b_\mathrm{UTP}$ should be greater than 1, as
was found for  
the ratio $a_\mathrm{dipole}/a_\mathrm{UTP}$ in the linear case, to further 
increase the speedup.

Recognizing that our pseudopotential gives accurate results with around 
$2000$-times smaller computational outlay than using the real dipolar 
interaction, we now investigate the third parameter that could affect the 
accuracy, interaction strength.

\begin{figure}[tb]
\includegraphics[width=\linewidth]{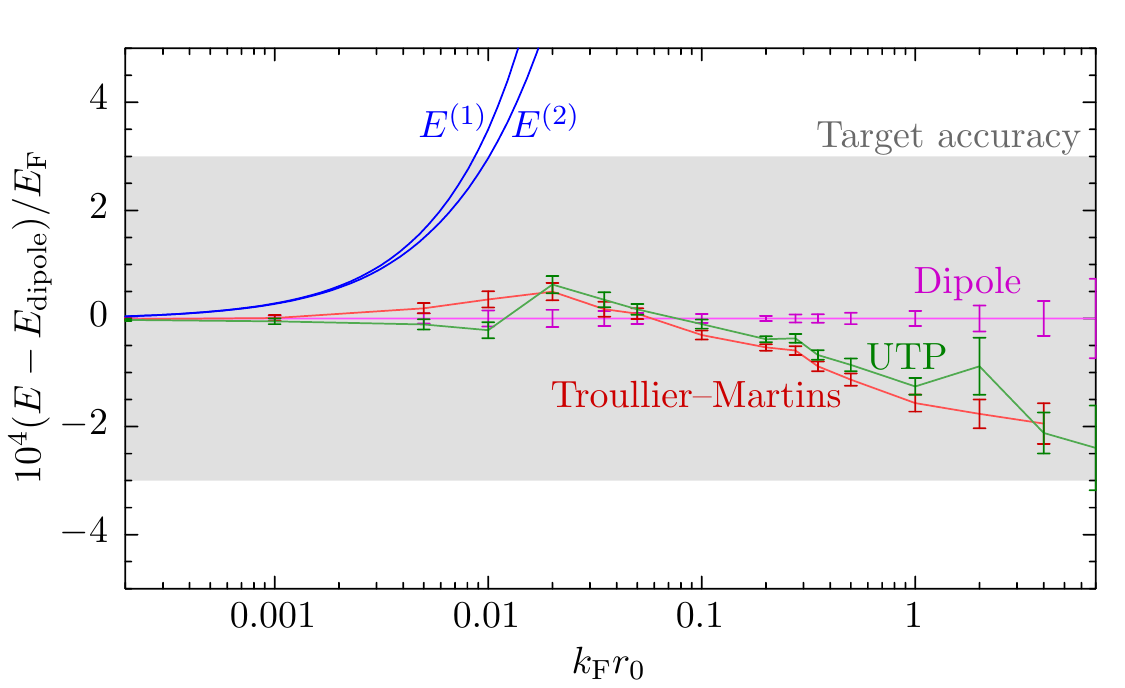}
\caption{(Color online) The deviation of the equation of state as calculated 
using the
  pseudopotentials from that calculated using the exact dipolar potential.  The 
dipolar potential is shown in magenta, with the Troullier--Martins 
pseudopotential in red, the UTP in green, and first- and second-order 
perturbation theory ($E^{(1)}$ and $E^{(2)}$) in blue.
  The gray box around the results using the dipolar potential shows the target
  $3\times 10^{-4}\Ef$ accuracy level.}
\label{dmc_high_energies}
\end{figure}

\subsection{Equation of state}
\label{results}

We compare the equations of state of the 2D dipolar Fermi gas as calculated
using the exact dipolar potential and the Troullier--Martins and UTP
pseudopotentials in \figref{dmc_energies}.  The pseudopotential cutoff is taken 
as $\rc=r_\mathrm{s}$ and we extrapolate to zero timestep following the 
procedure outlined in Reference~\cite{Lee11}.  We find the equations of state 
to be the same to order $10^{-4}\Ef$. Shown as black circles in 
\figref{dmc_energies} is the equation of state of the system as
calculated using DMC by Matveeva~and~Giorgini (MG) in 
Reference~\cite{Matveeva12}.  We explicitly repeat the simulation of 
Reference~\cite{Matveeva12}, using the same system of 81 particles, but our 
calculated energies using the dipolar potential are of order $10^{-2}\Ef$ lower 
than reported there, and as DMC is a variational technique this indicates that 
our trial wavefunction is likely more accurate than was available to the 
authors of Reference~\cite{Matveeva12}, possibly due to our inclusion of a 
Jastrow factor with variational parameters.  On the scale of 
\figref{dmc_energies} it is not possible to distinguish our pseudopotential 
calculations from those using the real dipolar interaction, and so in order to 
properly analyze them we examine the error from the true dipolar potential in 
\figref{dmc_high_energies}.

Following the accuracy used in Reference~\cite{Matveeva12} to draw conclusions 
about which phases are energetically favorable in the dipolar gas, we choose a 
target accuracy of $3\times 10^{-4}\Ef$ for our pseudopotentials, shown as a 
gray box in \figref{dmc_high_energies}.  Over a wide range of interaction 
strengths our pseudopotentials fall within this accuracy, with the UTP being 
slightly more accurate than the Troullier--Martins pseudopotential at most 
interaction strengths. We also compare our DMC results to second-order 
perturbation theory \cite{Lu12,Chan10}
\begin{align*}
E^{(2)}=\frac{\Ef}{2}
\left[1+\frac{128}{45\pi}\kf r_0+\frac{1}{4}(\kf r_0)^2\ln(1.43\kf r_0)\right],
\end{align*}
noting that it differs significantly from the DMC results above interaction 
strengths of $\kf r_0 \gtrsim 0.01$.  In \figref{dmc_energies} we also note 
that above $\kf r_0 \gtrsim 1$ first-order perturbation theory is more accurate 
than $E^{(2)}$, indicating that perturbation theory is not an adequate 
approximation except at very low interaction strengths $\kf r_0 \ll 0.01$.

\begin{figure}
\includegraphics[width=\linewidth]{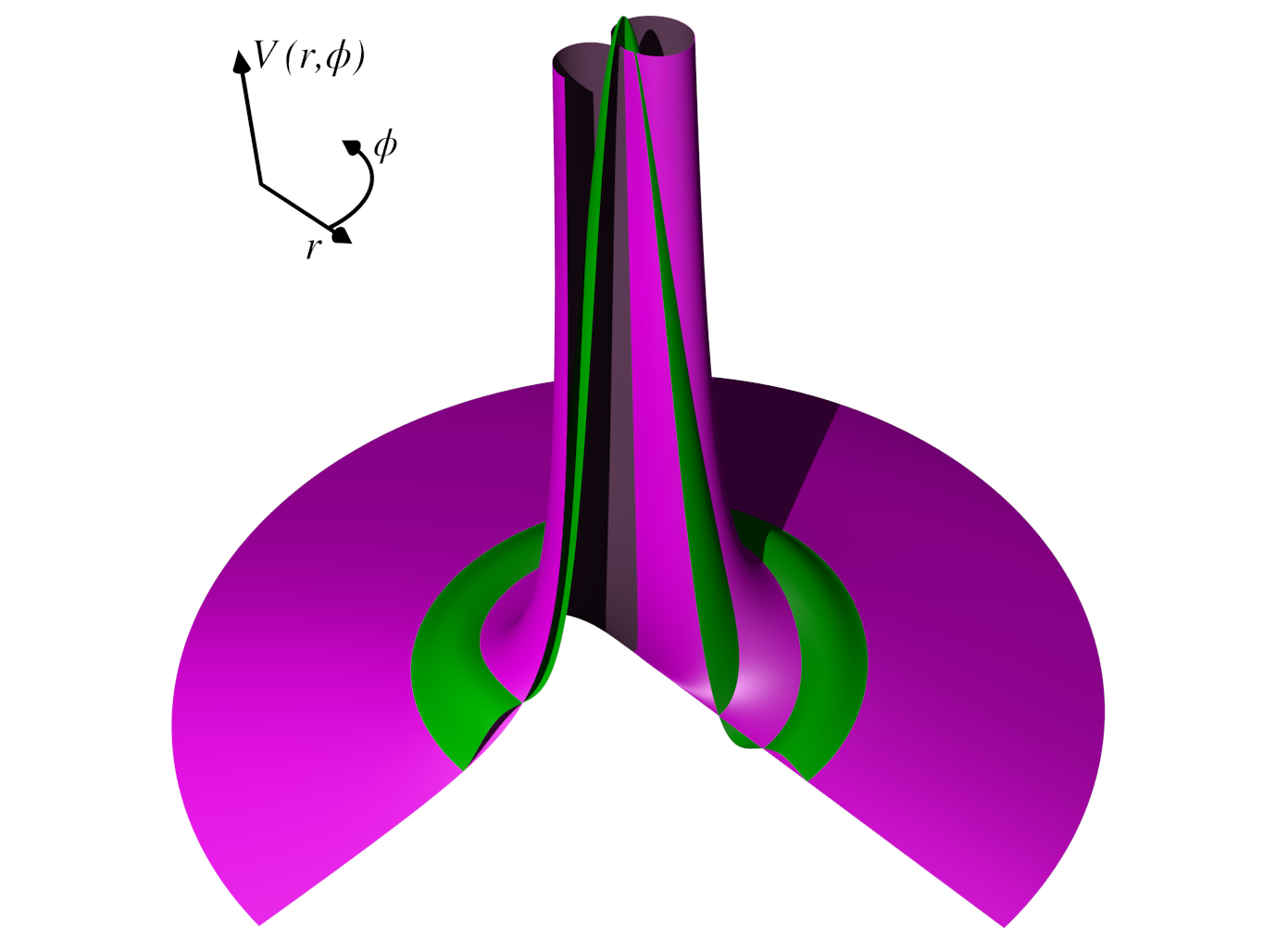}
\caption{(Color online) The dipolar potential $V(r,\phi)$ in magenta, and the 
UTP $V_\mathrm{UTP}(r,\phi)$ for the same tilt angle, in green.  The potentials 
are cut through for $3\pi/2<\phi<2\pi$ to contrast the radial variation of the 
dipolar potential along $\phi=3\pi/2$ and $\phi=2\pi$, and show the smooth join 
of the UTP onto the dipolar potential at $r=\rc$.}
\label{tilt_potential}
\end{figure}

We have constructed and tested pseudopotentials using the Troullier--Martins 
and UTP methods.  In each test, shown in 
Figures~\ref{phase_differences}(b),~\ref{harmonic_trap},~and~
\ref{dmc_high_energies}, the UTP method has given more accurate results.
  We therefore recommend 
the use of the UTP method to construct pseudopotentials for the dipolar 
interaction, and recommend its use over the dipolar potential with a cusp 
correction due to the 2000-times speedup in calculations that can be achieved 
whilst still achieving sufficient accuracy. We now go on to show that the UTP 
can be generalized to capture the effects of an anisotropic interaction in a 
system of tilted dipoles.

\section{Tilted dipoles}\label{sec:Tilted}

The above analysis has focused on dipoles aligned normal to their 2D plane of 
motion by an external electric or magnetic field.  However, this same electric 
or magnetic field could be used to align the dipoles at an angle $\theta$ to 
the normal to the plane \cite{Ni10}.  The dipolar interaction then takes the 
anisotropic form 
\mbox{$V(r,\phi)=d^2[1-\frac{3}{2}\sin^2\theta(1+\cos2\phi)]/r^3$} where $\phi$ 
is the polar angle in the plane, between the dipole-dipole separation and the 
projection of the electric field.  We focus on the $\theta \le 
\thetac=\mathrm{arcsin}(1/\sqrt{3})$ regime, where the potential is purely 
repulsive and there are no bound states. The potential $V(r,\phi)$ is shown in 
magenta in \figref{tilt_potential} for $\theta=\thetac$ and $\kf r_0=1/2$. As 
well as the $r^{-3}$ divergence, the potential is strongly anisotropic, 
separating into two lobes.  These properties make it difficult to work with 
numerically, and so we again develop a pseudopotential to ease the numerical 
simulation of this system.

The Troullier--Martins formalism used in the non-tilted system is not 
applicable to the case of $\theta>0$, and so here we propose the UTP

\begin{widetext}
\begin{align}
V_\mathrm{UTP}(r,\phi) = \frac{d^2}{\rc^3}
  \begin{cases} {\arraycolsep=1.4pt\def\arraystretch{2.2} \begin{array}{l}
    \left[1-\frac{3}{2}\sin^2\theta(1+\cos2\phi)\right] + 3\left( 
1-\frac{r}{\rc} \right) \left( \frac{r}{\rc} \right)^2 \left[ 1 - 
\frac{3}{2}\sin^2 \theta (1+\cos2\phi) \right]\\
\quad+ \left( 1 - \frac{r}{\rc} \right)^2 \left(1-\frac{3}{2} \sin^2\theta 
\right) \left[ v_1 \left( \frac{1}{2}+\frac{r}{\rc} \right) + 
\displaystyle\sum_{i=2}^{N_v} v_i \left(\frac{r}{\rc} \right)^i \right] \\
\quad + \sin^2 \theta \cos 2 \phi \left[ \left(1-\frac{r}{\rc}\right)^2 
v_{N_v+1} \left(\frac{r}{\rc}\right)^2  +3 \left( \frac{1}{2} - \frac{3}{2} 
\left(\frac{r}{\rc}\right)^2 + \left(\frac{r}{\rc}\right)^3 \right)  \right]
  \punc{,}
   \end{array} } & \begin{array}{c}\vspace{43pt}\\r<\rc\punc{,}\end{array}
  \vspace{8pt}\\
 \left[1-\frac{3}{2}\sin^2\theta(1+\cos2\phi)\right] \rc^3/r^3 \punc{,} & r \ge 
\rc\punc{,}
  \end{cases}
\label{TiltedPot}
\end{align}
\end{widetext}
which is constrained to be smooth to first derivative in both radial and 
azimuthal directions at the origin and at $\rc$, where it joins onto the exact 
dipolar potential.  $N_v$ is again set as $3$, and the coefficients $\{ v_i\}$ 
are minimized similarly to the non-tilted case.  At $\theta=0$ 
\eqnref{TiltedPot} reduces to the non-tilted form. A sample UTP is shown along 
with the tilted dipolar potential in \figref{tilt_potential}, demonstrating its 
non-divergent properties at particle coalescence and that it smoothly merges 
into the dipolar potential at $r=\rc$.  Furthermore, the angular variation of 
the UTP is less extreme than the real dipolar potential, which should lead to 
smoother estimates of the local energy at high tilt angles.

To optimize the pseudopotential we again calibrate in the two-body system.  The 
$\cos 2 \phi$ term in the potential couples together angular momentum channels 
of the wavefunction that differ by two angular momentum quanta, meaning that we 
can no longer solve the Schr\"odinger Equation separately in each angular 
momentum channel. Now that weight will be passed between the channels, they 
need to be considered explicitly and simultaneously. 

We solve the Schr\"odinger Equation simultaneously in the lowest four occupied 
angular momentum channels, $\ell=\{1,3,5,7\}$, numerically for both the dipolar 
potential and, separately, using the pseudopotential, in order to find values 
for the coefficients $\{ v_i\}$. As part of this process we optimise the weight 
in each channel. Unlike in the $\theta=0$ case it is not possible to find an 
analytic scattering wavefunction in the two-body homogeneous system that 
correctly captures the physics of the system in any limit.  Instead we optimise 
the parameters $\{ v_i\}$ by matching the energy of two particles in an 
harmonic trap, in effect minimising the error that was shown in 
\figref{harmonic_trap}.

\begin{figure}[tb]
\includegraphics[width=\linewidth]{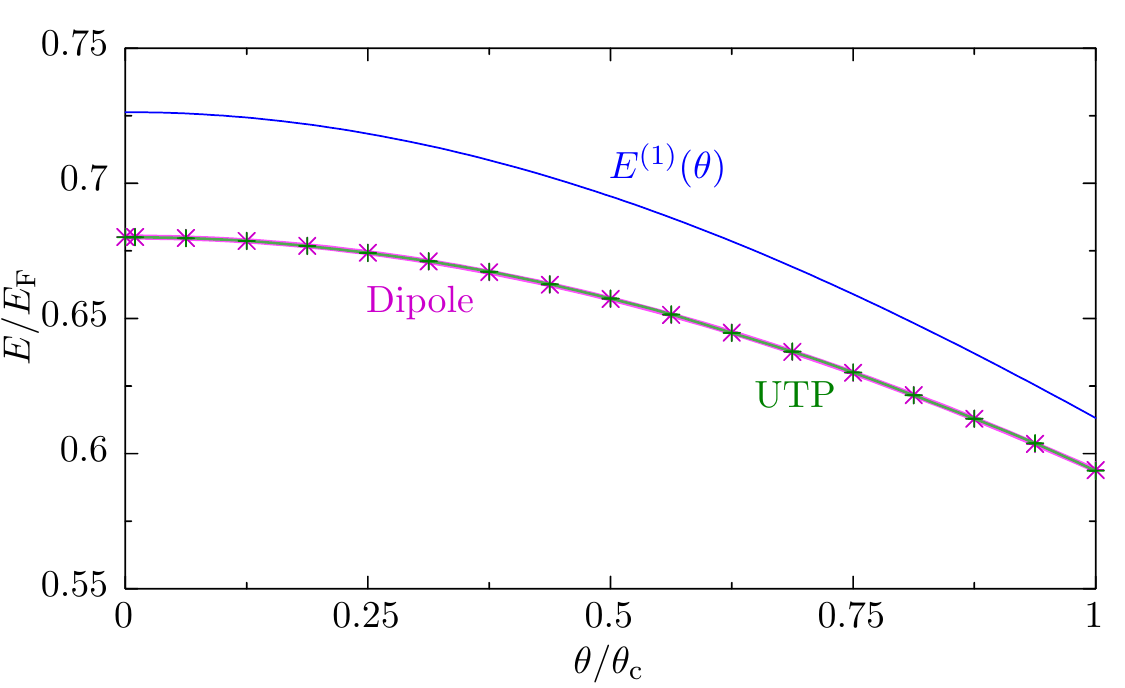}
\caption{(Color online) The equation of state of the tilted dipolar gas system 
as a function of tilt angle $\theta$.  Our DMC data using the dipolar potential 
and UTP overlie one another to within the width of the plotted lines, with 
stochastic error bars of order $10^{-5}\Ef$.  First-order perturbation theory 
$E^{(1)}$ is shown in blue.}
\label{tilt_energies}
\end{figure}

We need to select an optimal trap frequency $\omega$ at which to calibrate the 
pseudopotential.  To do this, we re-write the reduced system Hamiltonian for 
particles in an harmonic trap as $\hat{H}=\hat{H}_\mathrm{iso}(\hat{r}) + 
\hat{H}_\mathrm{aniso}(\hat{r},\hat{\phi})$, with
\begin{align*}
\hat{H}_\mathrm{iso}(\hat{r})&=-\nabla^2 + \frac{1}{4} \omega^2 \hat{r}^2 + 
\frac{\dbar^2}{\hat{r}^3}, \\
\hat{H}_\mathrm{aniso}(\hat{r},\hat{\phi})&=- 
\frac{\dbar^2}{\hat{r}^3}\frac{\frac{3}{2}\sin^2 \theta}{1-\frac{3}{2}\sin^2 
\theta} \cos 2 \hat{\phi},
\end{align*}
and $\dbar^2=d^2 (1-\frac{3}{2}\sin^2 \theta)$.  $\hat{H}_\mathrm{iso}$ 
captures the effect of the harmonic trap and the isotropic part of the dipolar 
interaction, whilst $\hat{H}_\mathrm{aniso}$ captures the anisotropic part of 
the dipolar interaction. We seek a trap frequency $\omega$ at which the average 
kinetic energy of the harmonic trap system is the same as that of the 
homogenous system, allowing us to select the appropriate Fermi momentum $\kf$ 
to describe the interaction strength $\kf r_0$.  For the isotropic part of the 
Hamiltonian we can apply a cusp correction to the non-interacting harmonic trap 
wavefunction, in the same spirit as \secref{KatoCusp}.  This gives a trial 
wavefunction
\begin{align*}
\psi(r,\phi)\propto \omega\, r\, \expe^{-\frac{1}{4}\omega 
r^2-2\frac{d}{\sqrt{r}}}.
\end{align*}
We set the average kinetic energy of the isotropic harmonic trap system as 
equal to the kinetic energy of the homogeneous system and solve for $\omega$, 
which for interaction strength $\kf r_0=1/2$ is $\omegaiso\approx 2.2\Ef$.

\begin{figure}[tb]
\includegraphics[width=\linewidth]{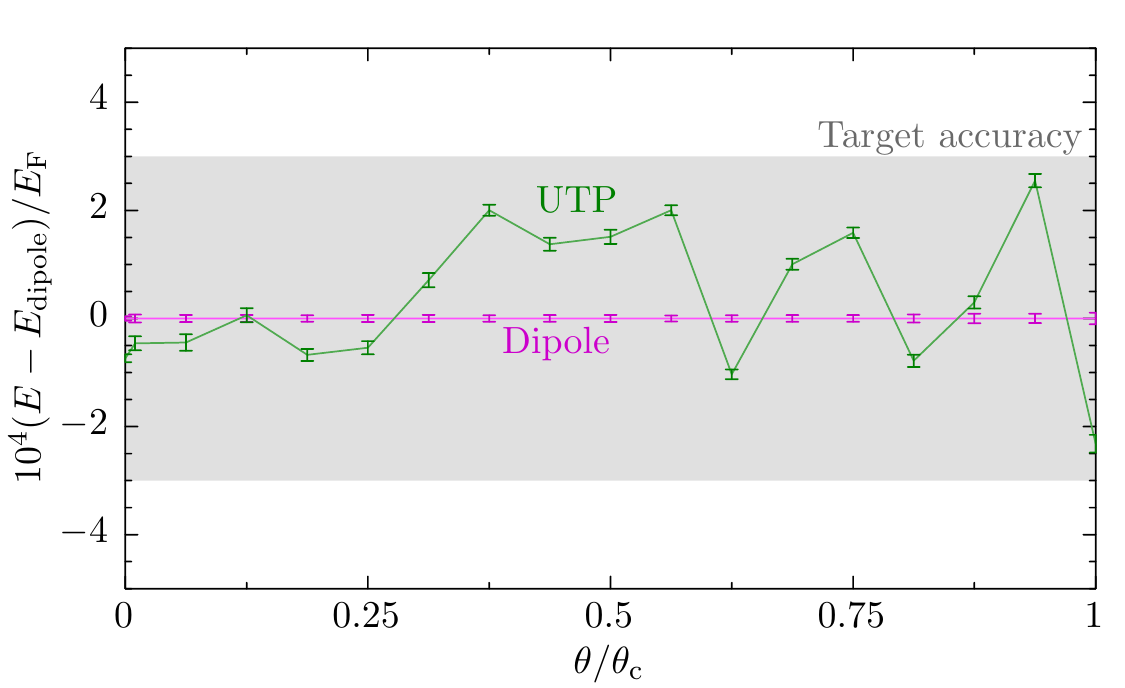}
\caption{(Color online) The deviation of the equation of state as calculated 
using the tilted
  pseudopotential from that calculated using the exact dipolar potential.  
Results using the dipolar potential are shown in magenta, with those using UTP 
pseudopotential in green.
  Similarly to \figref{dmc_high_energies}, the gray box around the results 
using the dipolar potential shows the targeted
  $3\times10^{-4}\Ef$ accuracy level.}
\label{high_tilt_energies}
\end{figure}

Having analyzed the isotropic part of the Hamiltonian we now turn to the 
anisotropic $\hat{H}_\mathrm{aniso}$.  As there is no analytical solution to 
the tilted two-body scattering problem available we instead perform a 
perturbative analysis in small $\theta$.  We search for the most important 
contribution that $\hat{H}_\mathrm{aniso}$ makes to the system's energy, which 
occurs where $|\psi(r,\phi)\hat{H}_\mathrm{aniso}\psi(r,\phi)|$ is maximal.  
This is at $r\approx r_0$ and $\phi=0$, and using these values in the 
functional form of $\hat{H}_\mathrm{aniso}$ we get a perturbative energy 
\mbox{$\frac{3}{2}r_0^{-2}\sin^2 \theta\left(1-\frac{9}{4}\sin^4\theta\right)$} 
for small $\theta$.  Adding this to the isotropic trap frequency we obtain the 
harmonic trap freqency \mbox{$\omega\approx 2.2\Ef + \frac{3}{2}r_0^{-2}\sin^2 
\theta\left(1-\frac{9}{4}\sin^4\theta\right)$}, which we use to optimise the 
pseudopotentials. An example UTP is shown in \figref{tilt_potential}, 
demonstrating its smooth and non-divergent properties.  The form of the 
pseudopotential is robust against changes in the trap frequency $\omega$ used 
to construct it.
With the pseudopotential in place we perform DMC calculations to evaluate the 
ground state energy of the anisotropic, homogeneous dipolar gas. In 
\figref{tilt_energies} we show the equation of state of the tilted dipole gas 
at interaction strength $\kf r_0=1/2$ over a range of tilt angles 
$0\le\theta\le\thetac$ away from vertical. We use a similar trial wavefunction 
to the non-tilted case, with the addition to the Jastrow factor of an 
anisotropic term
\begin{align*}
\prod_{i\neq j} \exp\left[\left( \sum_{k=0}^{N_\mathrm{s}} s_k r^k_{ij} \cos 
\left( 2 \phi_{ij}\right)\right) \left(1 - \frac{r_{ij}}{L}\right)^3 
\Theta(L-r_{ij})\right],
\end{align*}
where the variables have the same meaning as in \eqnref{Jastrow_factor}, 
$\phi_{ij}$ is the polar angle between the particles labelled $i,j$, and 
$N_\mathrm{s}=6$.  This term captures the leading-order anisotropies in the 
inter-particle correlations.  The addition of higher-order angular terms did 
not provide any significant benefit.  In calculations using the real tilted 
dipolar potential we also modify the cusp condition to the form $\prod_{i > 
j}\exp(-2 \dbar /\sqrt{r_{ij}})$.

In \figref{tilt_energies} we compare our DMC estimates of the equation of state 
to first-order perturbation theory \cite{Parish12}
\begin{align*}
E^{(1)}(\theta)=\frac{\Ef}{2}\left[ 1+\frac{128}{45\pi} \kf r_0 \left( 
1-\frac{3}{2}\sin^2 \theta \right) \right].
\end{align*}
Similarly to the non-tilted case we find that perturbation theory overestimates 
the energy, and also that it overestimates the reduction in energy with 
increasing tilt angle.  Again the results using the exact dipolar interaction 
and those using our UTP are so similar they cannot be distinguished on this 
scale, and so we analyze the pseudopotential accuracy by examining the energy 
error from the dipolar potential in \figref{high_tilt_energies}. As in the 
non-tilted system the pseudopotential achieves our target accuracy of $3\times 
10^{-4} \Ef$ across a wide range of parameter space. The pseudopotential is 
particularly accurate below $\theta\lesssim \thetac/4$ where there is less 
coupling between angular momentum channels, at $\theta\to 0$ reproducing the 
same accuracy that was found in the non-tilted system.

\begin{figure}[tb]
\includegraphics[width=\linewidth]{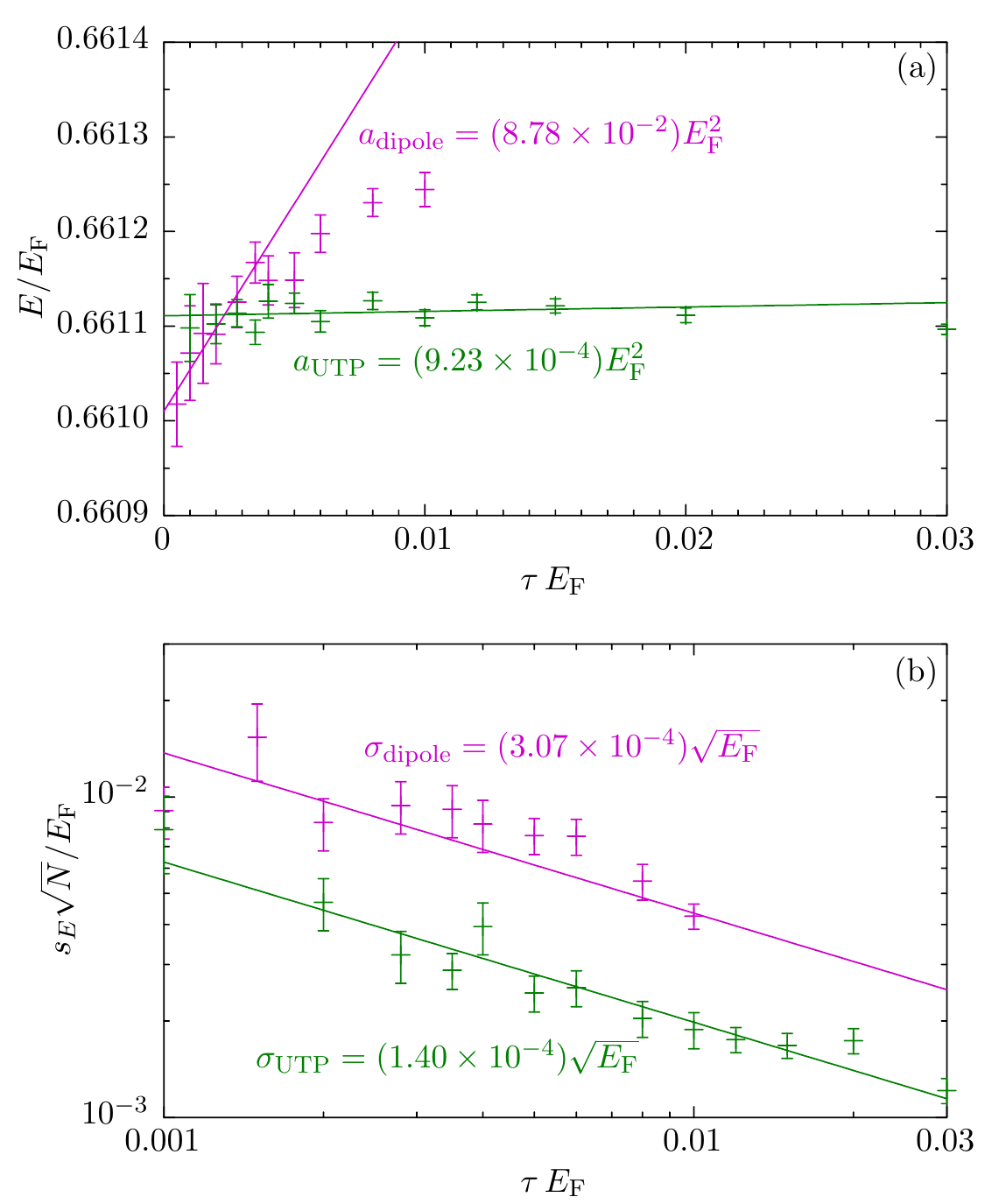}
\caption{(Color online) (a) The variation of the energy per particle in the 
Fermi gas of tilted dipoles with timestep $\tau$, with the values of the linear 
error parameters $a$. (b) The standard error $s_\mathrm{E}$ in the energy per 
particle in the Fermi gas, again with fitted $1/\tau$ parameters given.}
\label{tilt_dtdmc}
\end{figure}

To determine the full benefit of using the pseudopotential in a tilted system 
we examine the behavior of the calculated energy with DMC timestep in 
\figref{tilt_dtdmc}, evaluated at $\kf r_0=1/2$ and $\theta=\theta_c/2$.  
Similarly to the non-tilted case, \figref{tilt_dtdmc}(a) shows that the energy 
calculated using the pseudopotential has significantly improved behavior with 
timestep when compared to the dipolar potential, having less severe variation 
of the energy with timestep and also remaining in the linear regime out to 
larger $\tau$.  There is also a reduction in standard error of $\sim 2.2$ times 
when using the pseudopotential, as seen in \figref{tilt_dtdmc}(b). Combining 
the fitting parameters in \figref{tilt_dtdmc} in the way set out in 
\secref{formalism} shows the pseudopotential to be $\sim 450$ times quicker to 
use than the real tilted dipolar interaction.

We have constructed pseudopotentials for the dipolar interaction at tilt angles 
$0\le \theta \le \thetac$, and shown that they give the ground state energy of 
the anisotropic, homogeneous dipolar gas to within $3\times 10^{-4} \Ef$, and 
also provide a $450$-times speedup over using the real tilted dipolar 
interaction.  This means that they will be an accurate and efficacious tool to 
carry out DMC investigations of the whole $0\le \theta \le \thetac$ phase 
diagram.

\section{Discussion}\label{Discussion}

We have developed accurate pseudopotentials for the dipolar interaction
in two dimensions and tested them against the dipolar interaction by comparing 
scattering phase shifts, energies in an
harmonic trap, and the ground state of a Fermi gas.  The pseudopotentials 
deliver ground state energies of the Fermi gas to an accuracy of $3\times 
10^{-4}\Ef$, and their smoothness accelerates DMC calculations by a factor of 
up to $\sim2000$.

The pseudopotentials have been constructed to work in situations where the 
dipole moments are aligned both normal and at an angle to the two-dimensional 
plane of motion of the particles.  This could allow the formalism developed 
here to be used in an analysis of the full phase diagram of the 2D dipolar gas, 
including investigating the high interaction strength regime where the Fermi 
fluid forms a Wigner-type crystal \cite{Matveeva12}, possibly after passing 
through a stripe phase \cite{Parish12}, or to turn to the tilted section of the 
phase diagram, with the possibility of superfluid behavior at high tilt angles 
\cite{Bruun08}.  Superfluidity is also expected in a system of dipoles dressed 
by an external microwave field \cite{Cooper09,Levinsen11}, a system that would 
also be amenable to analysis using a pseudopotential.  The method used here for 
constructing pseudopotentials for the tilted system could also be extended to a 
3D system of dipolar particles, or to study a classical analogue of the system.

\acknowledgments{The authors thank Pascal Bugnion, Neil Drummond, Pablo
  L\'opez R\'ios, and Richard Needs for useful discussions. TMW acknowledges 
the financial support of the EPSRC [EP/J017639/1], and GJC
  acknowledges the financial support of the Royal Society and Gonville \&
  Caius College. There is Open Access to this paper and data available at 
\texttt{https://www.repository.cam.ac.uk}.}

\appendix

\section{Construction of the Troullier--Martins pseudopotentials}
\label{TroullierMartins}

The Troullier--Martins formalism is a method for developing pseudopotentials 
that were originally designed for use in electron-ion calculations 
\cite{Troullier91}.  Here, following Reference~\cite{Bugnion14} we adapt it to 
the case of a 2D dipolar potential. The scattering Schr\"odinger 
Equation~\eqref{SchrodingerEquation} may be
written in 2D circular coordinates $(r,\phi)$ as
\begin{align*}
&-\left( \frac{1}{r} \frac{\partial}{\partial r} \left( r 
\frac{\partial}{\partial r} \right) + \frac{1}{r^2} \frac{\partial^2}{\partial 
\phi^2} \right) \psi(r,\phi) + V(r)\psi(r,\phi) \neweqnline
&\qquad= E \psi(r,\phi)\punc{,}
\end{align*}
where we wish to replace the dipolar potential \mbox{$V(r) = d^2/r^3$} with
a pseudopotential inside a cutoff radius $\rc$.  Expanding the wavefunction in 
angular momentum channels as
\begin{align*}
\psi (r,\phi) = \sum_{\ell=0}^{\infty}  r^\ell \psi_\ell(r) \cos(\ell \phi)
\end{align*}
we obtain a radial equation for the wavefunction  $\psi_\ell$ in each channel
\begin{align}
-\left( \frac{2\ell+1}{r}\psi_\ell' + \psi_\ell'' \right) + V(r) \psi_\ell = E 
\psi_\ell\punc{,}
\label{Scheq}
\end{align}
where the primes indicate differentiation with respect to $r$.  We choose a
calibration energy $\Ec$ at which the pseudopotential will exactly replicate
the dipolar potential's scattering characteristics, whose optimal choice is 
found in
\appref{CalibrationEnergy} to be $\Ef/4$. We then
construct the pseudopotential by working from a pseudo-wavefunction that within 
a radius $\rc$ takes the form
\begin{align*}
\psi_{\mathrm{pseudo},\ell}(r) = \expe^{p(r)} \punc{,}
\end{align*}
where $p(r) = \sum_{i=0}^6 c_i r^{2i}$.  The form $\expe^{p(r)}$ is positive
definite, which ensures that no spurious nodes are introduced into the 
wavefunction.  Inserting the wavefunction into
\eqnref{Scheq} we find that the pseudopotential in each angular 
momentum channel $\ell$ should take the form
\begin{align}
V_{\mathrm{T\textendash M}}(r) = \left\{ \begin{array}{ll} \Ec + 
\frac{2\ell+1}{r} p' + p'^2 + p'', & r < \rc, \\                                           
     d^2/r^3, & r \ge \rc\punc{.}
\end{array} \right.
\label{pseudo}
\end{align}

In order to calculate $p(r)$ explicitly we impose a series of constraints
on it: firstly, that the pseudo-wavefunction's value and first four 
derivatives match those of the exact
wavefunction at $\rc$, in order that the first two derivatives of the 
pseudopotential are continuous,
\begin{align*}
p(\rc) &= \ln\left( \frac{R_\ell(\rc)}{\rc^{\ell+1}} \right), \\
p'(\rc) &= \frac{R_\ell'(\rc)}{R_\ell(\rc)} - \frac{\ell+1}{\rc}, \\
p''(\rc) &= V(\rc) - \Ec - (p'(\rc))^2 - \frac{2\ell+1}{\rc}p'(\rc), \\
p'''(\rc) &= V'(\rc) - 2 p'(\rc) p''(\rc) - \frac{2\ell+1}{\rc} p''(\rc) \\
&\quad+ \frac{2\ell+1}{\rc^2} p'(\rc), \\
p''''(\rc) &= V''(\rc) - 2 (p''(\rc))^2 - 2 p'(\rc) p'''(\rc)  \\
&\;- \frac{2\ell+1}{\rc} p'''(\rc) + 2\frac{2\ell+1}{\rc^2}p''(\rc) - 2 
\frac{2\ell+1}{\rc^3}p'(\rc),
\end{align*}
where $R_\ell(r) = r \psi_{\mathrm{dipole},\ell}(r)$.  The 
polynomial form of $p(r)$ ensures that this is a set of linear equations 
in the coefficients $c_i$, and so has a straighforward solution. 
We also require that the pseudo-wavefunction has zero curvature at the origin, 
\begin{align*}
c_2^2 &= - c_4 (2\ell+4),
\end{align*}
and that the norm of the pseudo-wavefunction within the cutoff radius is the 
same as that
from the exact potential, to conserve the physical particle weight
\begin{align*}
2 c_0 &+ \ln \left( \int_0^{\rc} r^{2\ell+1} \exp\left(2p(r) - 2 c_0\right) \dd
r \right) \\
&= \ln\left( \int_0^{\rc} |\psi_{\mathrm{dipole},\ell}(r,\phi)|^2 r \dd r 
\right).
\end{align*}
This fully specifies $p(r)$ and hence, via Equation~\eqref{pseudo}, 
$V_{\mathrm{T\textendash M}}$.  We solve these equations simultaneoulsy for the 
$c_i$, always taking the branch of the quadratic equation that gives the 
smaller value for $c_0$, which in turn gives a larger reduction in variance for 
simulations using the pseudopotential.

\section{Choosing a calibration energy}
\label{CalibrationEnergy}

The Troullier--Martins formalism for deriving pseudopotentials is designed to
give exact scattering properties at the calibration energy.  The
norm-conservation condition may also be considered as requiring that the
derivative of the phase shift with respect to energy evaluated at the
calibration energy $\partial \Delta \delta / \partial E |_{\Ec} = 0$
\cite{Bugnion14}.  This means that to leading order the error in the
scattering phase shift when using a Troullier--Martins pseudopotential $\Delta
\delta \propto (E - \Ec)^2$.  Expressing this in terms of the relative
momentum $\vect{k}_1 - \vect{k}_2$ of the two scattering particles with momenta 
$\vect{k}_1$, $\vect{k}_2$, the scattering phase shift error $\Delta 
\delta(\left|
    (\vect{k}_1 - \vect{k}_2)/2 \right|^2) \propto (
  \left|(\vect{k}_1 - \vect{k}_2)/2 \right|^2 - \kc^2 )^2$ where
$\kc=\sqrt{\Ec}$ is the calibration wave vector.  To find the optimum 
calibration wave
vector we average this error over the Fermi sea for particles 1 and 2 and
then minimize with respect to $\kc$. The average
\begin{align}
\langle \Delta \delta \rangle = \frac{\displaystyle\int \Delta 
\delta\left(\left|\frac{\vect{k}_1 - \vect{k}_2}{2}\right|^2\right) n(k_1) 
n(k_2)\, \dd \vect{k}_1\, \dd \vect{k}_2}{\displaystyle\int  n(k_1) n(k_2)\, 
\dd \vect{k}_1\, \dd \vect{k}_2},
\label{Delta_delta_k}
\end{align}
where $n(k)$ is the Fermi-Dirac distribution, can be re-written in terms of
center-of-momentum and relative coordinates $\vect{x}=(\vect{k}_1 -
\vect{k}_2)/2 k_{\mathrm{F}}$, $\vect{y}=(\vect{k}_1 + \vect{k}_2)/2
k_{\mathrm{F}}$, which transforms Equation~\eqref{Delta_delta_k} into
\cite{Lu12}
\begin{align*}
\langle \Delta \delta \rangle \propto \int_0^{2\pi} \int_0^1 \int_0^{y_0 
(x,\phi)} \Delta \delta ( k_{\mathrm{F}}^2 x^2)\, x \, y \, \dd y\, \dd x\,\dd 
\phi \punc{,}
\end{align*}
where $\phi$ is the angle between $\vect{x}$ and $\vect{y}$ and the function
$y_0 (x,\phi) = - x |\cos \phi| + \sqrt{1 - x^2 \sin^2 \phi} $.  This then
simplifies to
\begin{align*}
\langle \Delta \delta \rangle \propto\!\!\! \int_0^1\!\! x \Delta \delta ( 
\kf^2 x^2) \left( \pi - 2 \left( x \sqrt{1-x^2} + \arcsin x \right) \right) 
\,\dd x \punc{,}
\end{align*}
and substituting the form of 
$\Delta \delta\propto(\kf^2 x^2 - \kc^2)^2$ 
from above the optimum value of $\kc$
is found to be $\kf/2$, and hence the optimum calibration energy $\Ec =
\Ef/4$.

\end{document}